\documentclass[twocolumn]{aastex631}

\usepackage{longtable}
\usepackage[flushleft]{threeparttable}
\usepackage{tabularx}
\usepackage{multirow}
\usepackage{graphicx}
\usepackage{amsmath,amssymb}
\usepackage{color}
\usepackage{units}
\usepackage{epstopdf}
\usepackage{hyperref}
\usepackage{multirow}
\usepackage{url}
\usepackage{subfigure}
\usepackage{rotating}
\usepackage{enumitem}\setlist[description]{font=\textendash\enskip\scshape\bfseries}

\newcommand{\beq}{\begin{equation}}
\newcommand{\eeq}{\end{equation}}
\newcommand{\bdm}{\begin{displaymath}}
\newcommand{\edm}{\end{displaymath}}
\newcommand{\T}[1]{\tilde{#1}}

\definecolor{Gray}{gray}{0.9}
\definecolor{orange}{rgb}{0.9,0.5,0}

\newcommand{\ttt}[1]{\texttt{#1}}

\newcommand{\ztfink}[1]{ZTF-Fink}

\begin{document}

\title{A data science platform to enable time-domain astronomy}

\author[0000-0002-8262-2924]{Michael W. Coughlin}
\affil{School of Physics and Astronomy, University of Minnesota, Minneapolis, Minnesota 55455, USA}

\author[0000-0002-7777-216X]{Joshua S. Bloom}
\affil{Department of Astronomy, University of California,
Berkeley, CA 94720, USA}
\affil{Lawrence Berkeley National Laboratory, 1 Cyclotron Road,
MS 50B-4206, Berkeley, CA 94720, USA}

\author[0000-0002-7501-5579]{Guy Nir}
\affil{Department of Astronomy, University of California,
Berkeley, CA 94720, USA}
\affil{Lawrence Berkeley National Laboratory, 1 Cyclotron Road,
MS 50B-4206, Berkeley, CA 94720, USA}

\author[0000-0002-7686-3334]{Sarah Antier}
\affil{Artemis, Observatoire de la Côte d’Azur, Université Côte d’Azur, Boulevard de l'Observatoire, 06304 Nice, France}

\author[0009-0003-6181-4526]{Theophile Jegou du Laz}
\affil{Division of Physics, Mathematics, and Astronomy, California Institute of Technology, Pasadena, CA 91125, USA}

\author[0000-0001-9276-1891]{Stéfan van der Walt}
\affil{Berkeley Institute for Data Science, University of California Berkeley, Berkeley, CA 94720, USA}

\author[0000-0002-7183-0410]{Arien Crellin-Quick}
\affiliation{Weights and Biases, Inc., 1479 Folsom Street, San Francisco, CA 90063, USA}

\author[0009-0008-3603-0013]{Thomas Culino}
\affil{ESILV - École Supérieure d’Ingénieurs Léonard de Vinci, Paris, France}

\author[0000-0001-5060-8733]{Dmitry A. Duev}
\affiliation{Weights and Biases, Inc., 1479 Folsom Street, San Francisco, CA 90063, USA}

\author[0000-0003-3461-8661]{Daniel A. Goldstein}
\affiliation{Weights and Biases, Inc., 1479 Folsom Street, San Francisco, CA 90063, USA}

\author[0000-0002-7718-7884]{Brian F. Healy}
\affil{School of Physics and Astronomy, University of Minnesota, Minneapolis, Minnesota 55455, USA}

\author[0000-0003-2758-159X]{Viraj Karambelkar}
\affil{Division of Physics, Mathematics, and Astronomy, California Institute of Technology, Pasadena, CA 91125, USA}

\author[0009-0007-0764-7367]{Jada Lilleboe}
\affil{School of Physics and Astronomy, University of Minnesota, Minneapolis, Minnesota 55455, USA}

\author[0000-0002-1486-3582]{Kyung Min Shin}
\affiliation{EnergyHub, Inc., 41 Flatbush Ave, Suite 400A, Brooklyn, NY 11217, USA}

\author[0000-0001-9898-5597]{Leo P. Singer}
\affil{Astroparticle Physics Laboratory, NASA Goddard Space Flight Center, Code 661, Greenbelt, MD 20771, USA}

\author[0000-0002-2184-6430]{Tom{\'a}s Ahumada}
\affil{Division of Physics, Mathematics, and Astronomy, California Institute of Technology, Pasadena, CA 91125, USA}

\author[0000-0003-3768-7515]{Shreya Anand}
\affil{Division of Physics, Mathematics, and Astronomy, California Institute of Technology, Pasadena, CA 91125, USA}

\author[0000-0001-8018-5348]{Eric C. Bellm}
\affil{DIRAC Institute, Department of Astronomy, University of Washington, 3910 15th Avenue NE, Seattle, WA 98195, USA}

\author[0000-0002-5884-7867]{Richard Dekany}
\affil{Caltech Optical Observatories, California Institute of Technology, Pasadena, CA 91125, USA}

\author[0000-0002-3168-0139]{Matthew J. Graham}
\affil{Division of Physics, Mathematics, and Astronomy, California Institute of Technology, Pasadena, CA 91125, USA}

\author[0000-0002-5619-4938]{Mansi M. Kasliwal}
\affil{Division of Physics, Mathematics, and Astronomy, California Institute of Technology, Pasadena, CA 91125, USA}

\author[0009-0003-6260-2891]{Ivona Kostadinova}
\affil{Division of Physics, Mathematics, and Astronomy, California Institute of Technology, Pasadena, CA 91125, USA}

\author[0000-0002-9108-5059]{R. Weizmann Kiendrebeogo}
\affil{School of Physics and Astronomy, University of Minnesota, Minneapolis, Minnesota 55455, USA}
\affil{Artemis, Observatoire de la Côte d’Azur, Université Côte d’Azur, Boulevard de l'Observatoire, 06304 Nice, France}
\affil{Laboratoire de Physique et de Chimie de l’Environnement, Université Joseph KI-ZERBO, Ouagadougou, Burkina Faso}

\author[0000-0001-5390-8563]{Shrinivas R. Kulkarni}
\affil{Owens Valley Radio Observatory 249-17, California Institute of Technology, Pasadena, CA 91125, USA}

\author[0000-0001-9827-1463]{Sydney Jenkins}
\affil{Department of Physics, Massachusetts Institute of Technology, 77 Massachusetts Ave., Cambridge, MA 02139, USA}

\author[0000-0002-2249-0595]{Natalie LeBaron}
\affil{Department of Astronomy, University of California,
Berkeley, CA 94720, USA}

\author[0000-0003-2242-0244]{Ashish~A.~Mahabal}
\affil{Division of Physics, Mathematics and Astronomy, California Institute of Technology, Pasadena, CA 91125, USA}
\affil{Center for Data Driven Discovery, California Institute of Technology, Pasadena, CA 91125, USA}

\author[0000-0002-0466-1119]{James D. Neill}
\affil{Division of Physics, Mathematics, and Astronomy, California Institute of Technology, Pasadena, CA 91125, USA}

\author[0000-0002-3155-0385]{B. Parazin}
\affil{School of Physics and Astronomy, University of Minnesota, Minneapolis, Minnesota 55455, USA}
\affil{Northeastern University, Boston, MA 02115, USA}

\author[0000-0002-8560-4449]{Julien Peloton}
\affil{IJCLab, Univ Paris-Saclay, CNRS/IN2P3, Orsay, France}

\author[0000-0001-8472-1996]{Daniel A. Perley}
\affil{Astrophysics Research Institute, Liverpool John Moores University, \\ IC2, Liverpool Science Park, 146 Brownlow Hill, Liverpool L3 5RF, UK}

\author[0000-0002-0387-370X]{Reed Riddle}
\affil{Caltech Optical Observatories, California Institute of Technology, Pasadena, CA 91125, USA}

\author[0000-0001-7648-4142]{Ben Rusholme}
\affiliation{IPAC, California Institute of Technology, 1200 E. California Blvd, Pasadena, CA 91125, USA}

\author[0000-0002-2412-9728]{Jakob van Santen}
\affil{Deutsches Elektronen-Synchrotron DESY, Platanenallee 6, 15738 Zeuthen, Germany}

\author[0000-0003-1546-6615]{Jesper Sollerman}
\affil{The Oskar Klein Centre, Department of Astronomy, Stockholm University, AlbaNova, SE-10691, Stockholm, Sweden}

\author[0000-0003-2434-0387]{Robert Stein}
\affil{Division of Physics, Mathematics, and Astronomy, California Institute of Technology, Pasadena, CA 91125, USA}

\author[0000-0003-1835-1522]{D.~Turpin}
\affil{Universit\'e Paris-Saclay, Universit\'e Paris Cit\'e, CEA, CNRS, AIM, 91191, Gif-sur-Yvette, France}

\author[0000-0002-9998-6732]{Avery Wold}
\affiliation{IPAC, California Institute of Technology, 1200 E. California Blvd, Pasadena, CA 91125, USA}

\author[0009-0008-6162-7882]{Carla Amat}
\affil{ESILV - École Supérieure d’Ingénieurs Léonard de Vinci, Paris, France}

\author[0009-0005-4020-4483]{Adrien Bonnefon}
\affil{ESILV - École Supérieure d’Ingénieurs Léonard de Vinci, Paris, France}

\author[0009-0009-9174-0592]{Adrien Bonnefoy}
\affil{ESILV - École Supérieure d’Ingénieurs Léonard de Vinci, Paris, France}

\author[0009-0005-0586-7281]{Manon Flament}
\affil{ESILV - École Supérieure d’Ingénieurs Léonard de Vinci, Paris, France}

\author[0009-0000-4242-7406]{Frank Kerkow}
\affil{School of Physics and Astronomy, University of Minnesota, Minneapolis, Minnesota 55455, USA}

\author[0009-0001-7793-3680]{Sulekha Kishore}
\affil{Division of Physics, Mathematics, and Astronomy, California Institute of Technology, Pasadena, CA 91125, USA}

\author[0009-0009-6530-4295]{Shloke Jani}
\affil{School of Physics and Astronomy, University of Minnesota, Minneapolis, Minnesota 55455, USA}

\author[0009-0003-9879-7694]{Stephen K. Mahanty}
\affil{School of Physics and Astronomy, University of Minnesota, Minneapolis, Minnesota 55455, USA}

\author[0009-0001-9218-7196]{Céline Liu}
\affil{ESILV - École Supérieure d’Ingénieurs Léonard de Vinci, Paris, France}

\author[0009-0008-8840-1576]{Laura Llinares}
\affil{ESILV - École Supérieure d’Ingénieurs Léonard de Vinci, Paris, France}

\author[0009-0008-9749-4604]{Jolyane Makarison}
\affil{ESILV - École Supérieure d’Ingénieurs Léonard de Vinci, Paris, France}

\author[0009-0000-1435-9178]{Alix Olliéric}
\affil{ESILV - École Supérieure d’Ingénieurs Léonard de Vinci, Paris, France}

\author[0009-0008-8448-4773]{Inès Perez}
\affil{ESILV - École Supérieure d’Ingénieurs Léonard de Vinci, Paris, France}

\author[0009-0000-2196-8481]{Lydie Pont}
\affil{ESILV - École Supérieure d’Ingénieurs Léonard de Vinci, Paris, France}

\author[0009-0006-8436-3192]{Vyom Sharma}
\affil{School of Physics and Astronomy, University of Minnesota, Minneapolis, Minnesota 55455, USA}

\begin{abstract}
\texttt{SkyPortal} is an open-source software package designed to efficiently discover interesting transients, manage follow-up, perform characterization, and visualize the results. By enabling fast access to archival and catalog data, cross-matching heterogeneous data streams, and the triggering and monitoring of on-demand observations for further characterization, a \texttt{SkyPortal}-based platform has been operating at scale for $>$2\,yr for the Zwicky Transient Facility Phase II community, with hundreds of users, containing tens of millions of time-domain sources, interacting with dozens of telescopes, and enabling community reporting. While \texttt{SkyPortal} emphasizes rich user experiences (UX) across common frontend workflows, recognizing that scientific inquiry is increasingly performed programmatically, \texttt{SkyPortal} also surfaces an extensive and well-documented API system. From backend and frontend software to data science analysis tools and visualization frameworks, the \texttt{SkyPortal} design emphasizes the re-use and leveraging of best-in-class approaches, with a strong extensibility ethos. For instance, \texttt{SkyPortal} now leverages ChatGPT large-language models (LLMs) to automatically generate and surface source-level human-readable summaries. With the imminent re-start of the next-generation of gravitational wave detectors,  \texttt{SkyPortal} now also includes dedicated multi-messenger features addressing the requirements of rapid multi-messenger follow-up: multi-telescope management, team/group organizing interfaces, and cross-matching of multi-messenger data streams with time-domain optical surveys, with interfaces sufficiently intuitive for the newcomers to the field. This paper focuses on the detailed implementations, capabilities, and early science results that establishes \texttt{SkyPortal} as a community software package ready to take on the data science challenges and opportunities presented by this next chapter in the multi-messenger era. 
\end{abstract}



\section{Introduction}

The proliferation of data from synoptic sky surveys affords an ever-increasing discovery potential in time-domain astrophysics, on solar system, galactic, and extragalactic scales. However, for this potential to be fully and efficiently realized, the volume, velocity, and heterogeneity of such data must be managed, curated, and coordinated. And, as often discovery requires more data for novel insights, new sky events demand an optimization of followup observations across a growing array of precious and loosely federated resources.  We now recognize the need to change how astronomers analyze these data sets, with the application of data science principles to facilitate discovery and new insights.

Multi-messenger astronomy, with the integration of data sets from e.g., Advanced LIGO \citep{aLIGO} and Advanced Virgo \citep{adVirgo} for gravitational waves, e.g., IceCube \citep{AaAc2017} for neutrinos, e.g., the Zwicky Transient Facility (ZTF) \citep{Bellm:19:ZTFScheduler,Graham2018,DeSm2018,Masci2019} and the forthcoming Vera C. Rubin Observatory \citep{Ivezic2019} for optical astronomy, and $\gamma$-ray survey instruments such as the {\it Neil Gehrels Swift Observatory} mission \citep{GeCh2004} and Fermi's Gamma-ray Burst Monitor (Fermi-GBM) \citep{MeLi2009}, is unique and demanding for the number of integrated sources of data it requires to study targeted systems. As its quintessential example, the detection of GW\,170817 \citep{AbEA2017b} and its associated electromagnetic transients AT2017gfo \citep{CoFo2017,SmCh2017,AbEA2017f} and GRB\,170817A~\citep{GoVe2017,SaFe2017,AbEA2017e} have demonstrated the power of multi-messenger astronomy for measurements of the Hubble constant \citep{CoDi2019,CoAn2020,2017Natur.551...85A,HoNa2018,DiCo2020}, the neutron star equation of state, and $r$-process nucleosynthesis \citep{ChBe2017,2017Sci...358.1556C, CoBe2017,PiDa2017,RoFe2017,RaSo2018,SmCh2017,WaHa2019,KaKa2019} amongst many other science cases.

Previously, the large sky area of gravitational-wave localizations made searching for (and doing science with) counterparts particularly challenging \citep[e.g.,][]{Andreoni2019S190510g, CoAh2019b, Goldstein2019S190426c, Gomez2019, LuPa2019, AnCo2020, Ackley2020, Andreoni2020S190814bv, AnAg2020,GoCu2020,KaAn2020}. While localization accuracies  should improve somewhat \citep{PeSi2021} in the fourth LIGO-Virgo-KAGRA observing run (O4; \citealt{Co2020}), continued innovation in how searches are conducted remain paramount. To this end, we have undertaken an effort to integrate the capabilities from two ``Marshals,'' so named as software stacks designed to marshal candidates to follow-up telescopes, used during the third Observing run (O3): the Global Relay of Observatories Watching Transients Happen (GROWTH) \texttt{Target of Opportunity Marshal} \citep{CoAh2019} and the Global Rapid Advanced Network Devoted to the Multi-messenger Addicts (GRANDMA) Interface and
Communication for Addicts of the Rapid follow-up in the multi-messenger Era (iCARE) platforms \citep{Antier:2019pzz}, within \texttt{SkyPortal}, the open-source Target and Observation Manager (TOM) currently employed in ZTF Phase II \citep{WaCr2019}. These TOMs and Marshals, whose role in the time-domain ecosystem will be discussed below, are designed to support the astronomical community to efficiently work with these data sets, identify sources of interest, and obtain follow-up. In the following, we will use ``TOM'' to describe systems of this type, as opposed to ``Marshal,'' which can refer to systems more narrowly focused on follow-up, but acknowledge that these two terms are nearly synonymous in the community.

In this paper, we describe the components built within \texttt{SkyPortal}, focusing on their application to multi-messenger astronomy.
There are two existing platforms which deploy \texttt{SkyPortal} that heavily rely on these features: (i) \texttt{fritz}\footnote{\url{https://github.com/fritz-marshal/fritz}}: the platform used by the ZTF collaboration for Phase II operations and (ii)  \texttt{icare}\footnote{\url{https://github.com/grandma-collaboration/icare}}: the platform used by the GRANDMA collaboration.
These are private instances deploying \texttt{SkyPortal} for use by members of their collaborations; we will point out aspects that are deployment specific below to give readers a sense of what aspects may be deployment specific.
We describe our perspective on the time-domain ecosystem and how \texttt{SkyPortal} fits in it in Sec.~\ref{sec:ecosystem}.
The key features and technical implementation of the software stack are discussed in Sec.~\ref{sec:pipeline}.
Descriptions of the \texttt{fritz} and \texttt{icare} deployments, as well as an example analysis applied to ongoing $\gamma$-ray burst (GRB) searches are shown in Sec.~\ref{sec:science}.
We summarize our conclusions and future outlook, including our perspectives on software development in an academic environment, in Sec.~\ref{sec:conclusion}.

\section{The Time-Domain Software Ecosystem}
\label{sec:ecosystem}

\begin{figure*}[t]
    \centering
    \includegraphics[width=6.5in]{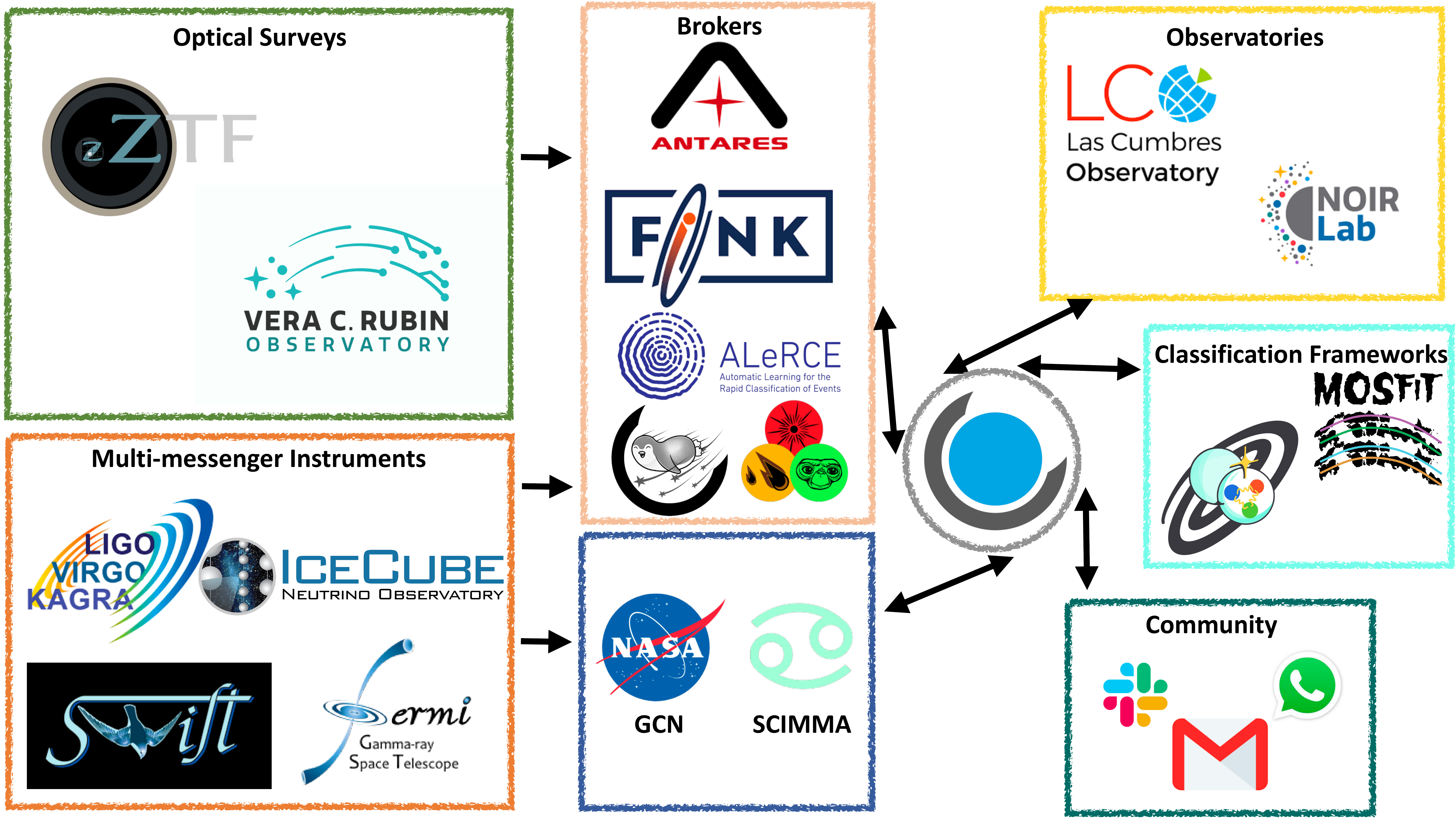}
    \caption{Flowchart for the multi-messenger, technical ecosystem envisioned in this platform.}
    \label{fig:flowchart}
\end{figure*}

We briefly review the time-domain software ecosystem here, and encourage interested readers to explore the references to understand this continually evolving ecosystem.
Understanding this ecosystem is required for understanding the role of \texttt{SkyPortal}; our vision for the role this software stack in this ecosystem is presented by Figure~\ref{fig:flowchart}. 
The flowchart indicates the inter-related chains of research infrastructure, both hardware and software, within which \texttt{SkyPortal} functions.

There are a number of other related software stacks available with some overlap with \texttt{SkyPortal}'s capabilities. For example, \texttt{YSE-PZ} is a platform with data-broker-like querying and filtering abilities, and that offers a turn-key solution to teams looking for a focus on transient survey management \citep{CoJo2023}. The Target and Observation Manager (TOM) Toolkit focuses on customization for different science use cases, deployable both as a stand-alone package as well as with its functionality integrated into larger ecosystems \citep{StBo2018}. \texttt{Astro-COLIBRI} is an API-based platform, accessible via a public website or through mobile apps, which provides a user-friendly overview of transient astronomical events \citep{ReSe2021}. \texttt{Astro-COLIBRI} enables users to quickly identify possible multi-wavelength or multi-messenger counterparts, and assess potential ground-based follow-up options. 

In some ways, \texttt{SkyPortal} has ambitions to be a sophisticated ``full-stack transient ecosystem'' that goes beyond a TOM of these kinds---integrating capabilities of telescope scheduling, observing optimization, spatial catalogs, astronomical data analysis, and others (see Sec.~\ref{sec:pipeline}) into a single software stack.

The first key piece of the software ecosystem are brokers like Fink \citep{MoPe2020} or ALerCE \citep{FoCa2021}, which ingest alerts from optical surveys such as ZTF \citep{Bellm:19:ZTFScheduler, Graham2018} or Rubin Observatory \citep{Ivezic2019} and provide value-added annotations such as cross-matching with other catalogs. A purpose of a broker is to act as a layer between alert issuers and the scientific community analyzing the alert data. It exposes one or more services to help the scientists to efficiently (manually and programmatically) analyze the alert data from telescopes and surveys. Generally, brokers collect and store alert data, enrich them with information from other surveys and catalogs or user-defined added values such as machine-learning classification scores, and redistribute the most promising events for further analyses, including follow-up observations. Large-scale surveys such as ZTF and Rubin Observatory have required broker teams to design and implement new technological approaches to operate in real time on large computing infrastructures for a wide variety of science cases. Recently, Rubin Observatory announced which Community brokers will have unrestricted access to the complete alert stream for the next decade\footnote{\url{https://www.lsst.org/scientists/alert-brokers}}: ALeRCE \citep{FoCa2021}, AMPEL \citep{Nordin:2019kxt}, ANTARES \citep{MaSt2021}, BABAMUL, Fink \citep{MoPe2020}, Lasair \citep{SmWi2019}, and Pitt-Google.  Most of these brokers are already operating on the ZTF alert stream, which has constituted a prime opportunity to engage projects with the scientific community while preparing the ground for the upcoming Rubin Observatory alert data. 

Some of these alert brokers also process alert data from other surveys and telescopes as well, including data at different wavelengths and from different messengers, and interested readers are encouraged to explore the references above for specifics. 
These annotations may include information from other wavelengths, spectroscopy, or historical data; they also often provide classification information, such as using machine learning classifiers that account for these annotations. All of these innovations are designed to support astronomers in purifying the alert streams to restrict the objects to those of scientific interest for particular users.

However, the role of a broker is not to ingest all the data that a user of such a broker might need for a complete analysis, as it is sometimes beyond the expertise or the roadmap of the broker team, or it would require a substantial amount of work to integrate, or simply because some data requires proprietary access. Instead, broker systems expose interoperable tools so that users can export enriched data elsewhere to continue the analysis. In practice, this is where TOMs and Marshals, for which interfaces with most brokers have been developed, come in, where the coordination of follow-up observations and further scientific analyses will take place after brokers enrich the data. 

The second key piece of the software ecosystem relates to platforms whereby multi-messenger instruments (e.g., high energy, gravitational waves, neutrinos, and transient phenomena in general) can share alerts and rapid communications; these currently include the General Coordinates Network (GCN; \citealt{SiRa2023}) and the Scalable Cyberinfrastructure to support Multi-Messenger Astrophysics\footnote{\url{https://scimma.org/}} (SCiMMA) project. These projects serve to ingest and then re-distribute these alerts to subscribing astronomers in the community. In this sense they are similar to brokers, although in general do not seek to annotate the ingested data streams, as the brokers do; it is again TOMs and Marshals that are often ingesting these streams, although some brokers also enable alert stream cross-matching with these multi-messenger notices.

TOM systems interact with both brokers and these multi-messenger alert distribution systems. While there is inevitably some redundancy with brokers especially, the focus of TOMs is to be a centralized database whereby coordination of follow-up observations of identified candidates (such as those passing the filters run within brokers) and support of astronomer science (such as integrating classification frameworks to be used on the resulting data sets) is possible. These TOM systems involve both human and algorithmic interactions, such as for the prioritization and submission of follow-up requests for target lists, as well as software analysis interfaces to light curve fitting codes like MOSFiT \citep{GuNi2018} or the ``Nuclear Physics - Multi-messenger Astronomy'' (NMMA) framework \citep{PaDi2022}.

TOM's provide an important conduit for information flow within this ecosystem. While many of these services are developed independently, they rely on well-defined interfaces such as Application Programming Interfaces (APIs) to interact. For time-domain astronomy in particular, they must also interact with flexible algorithmic frameworks which can plot and fit transients such as supernovae or tidal disruption events, as well as periodic and quasi-periodic objects such as eclipsing binaries. They also need to interact with core external services to different astronomical communities such as the Transient Name Server\footnote{\url{https://www.wis-tns.org/}} (TNS) or the Minor Planet Center\footnote{\url{https://minorplanetcenter.net/}} (MPC). While it is possible to downselect the services a TOM interacts with, it is inevitable that these communities overlap, given that, for example, transient astronomers are always interested in avoiding following up asteroids like those reported to the MPC.

Underlying the software components of the time-domain ecosystem are the methods by which data are stored, normalized, and exchanged. While image and tabular data are well-served by international (IVOA\footnote{https://www.ivoa.net}) standards
for accessing, querying, and exchange, the only aspect of the time domain ecosystem that is currently described is the reporting of astronomical transients (VOEvents; \citealt{2011ivoa.spec.0711S}). There is as yet no agreed data model, query protocol, or exchange format for timeseries\footnote{A timeseries exchange protocol, VOTimeseries, was proposed \citep{2008AN....329..284B} but never adopted by the IVOA.} or the semantically rich heterogeneous data items of annotations (human comments, classifications, etc.). The semi-structured nature of the latter of these means as well that they are difficult to map into existing regular data standards. \texttt{SkyPortal} manages heterogeneous input by enforcing data exchange with a strict API schema which can be accessed by any programmatic tool (and could be described as a VOSI interface [\citealt{2017ivoa.spec.0524G}]
for compliance). However, there is clearly a strong need within the burgeoning MMA community for a reengagement in defining schema, formats, and services for multi-messenger data exchange.  

\section{\texttt{SkyPortal} Framework}
\label{sec:pipeline}

In the following subsections, we present the key features and specific, implemented subsystems of \texttt{SkyPortal} in use to enable applications in time-domain astronomy. These features are available in every \texttt{SkyPortal} instance, upon download and deployment of the software stack.

\subsection{Associated software packages}

By design, \texttt{SkyPortal} is open source, and through its development, changes have been contributed to open-source, upstream packages such as \texttt{sncosmo} \citep{BaBa2016}; while it would be out-of-scope to describe all of the packages on which \texttt{SkyPortal} relies, we briefly describe two such packages, Baselayer and HEALPix-Alchemy \citep{SiPa2022}, for which development is particularly coordinated.

 \subsubsection{Baselayer}

The open-source package \texttt{baselayer}\footnote{\url{https://github.com/cesium-ml/baselayer}} provides the backbone of a generic science-application web framework on top of which \texttt{SkyPortal} is built. Originally built for the open-source time-domain machine learning application \texttt{cesium} \citep{2016arXiv160904504N}, \texttt{baselayer} implements, e.g., web socket communication between the backend and frontend, authentication, access control, microservices, configuration, process management, cron jobs, migrations, and external logging. As far as possible, it combines standard solutions, such as \texttt{Tornado}, \texttt{Nginx}, \texttt{supervisor}, and \texttt{Python Social Auth}, instead of implementing them from scratch. By delegating web infrastructure to \texttt{baselayer}, concerns are better separated and the \texttt{SkyPortal} team can focus on domain-specific concerns. That said, the \texttt{SkyPortal} team also contributes platform improvements upstream as required, exposing these to a wider audience. 

At the core of \texttt{baselayer}, we find \texttt{sqlalchemy}\footnote{\url{https://www.sqlalchemy.org/}}, an Object Relational Mapper (ORM) written in Python that allows easy interaction with a relational database, here being \texttt{PostgreSQL}\footnote{\url{https://www.postgresql.org/}} \citep{PostgreSQL}. Several base models---which are the abstractions of tables in the database---are defined to introduce the concept of users, permissions, roles, API tokens, and groups. An addon of \texttt{Python Social Auth} provides the necessary models for user authentication using many \texttt{openid} authentication services. In both \texttt{baselayer} and \texttt{SkyPortal}, authentication through Google services is the method of choice. \texttt{sqlalchemy} also allows handling multiple parallel and separate connections to the database named sessions. This is an important requirement for any API-based application, where multiple users need to read/write/modify/delete from the database simultaneously. Wherever interacting with the database is necessary, we open a session, add/modify/delete from that session, and then commit the changes. Changes are made to the database only once the commit method is called, and \texttt{PostgreSQL} is capable of resolving conflicts between multiple commit sessions if necessary.

For the users to interact with the API-based application that is \texttt{baselayer}, we define API Handlers, which are simply python classes with the standard GET, POST, DELETE, PUT, and PATCH method an API is expected to have. Then, \texttt{Tornado} maps each handler to the associated API endpoint. It is capable of calling the right API method of the handler and passing in-path parameters for each API call. In \texttt{SkyPortal}, everytime a new feature is needed, database models and handlers to interact with them can be added to the application following the same process. When it comes to permissions, \texttt{baselayer} provides python decorators that can be specified above each API method to specify what permissions or roles a user needs to use it. Similarly, database-level access restrictions can be defined in every \texttt {sqlalchemy} model, to ensure that users only have access to the data points that have been shared with them or with the groups they belong to.

Using a microservice architecture managed with \texttt{supervisord}, we can run multiple instances of the application in parallel to ensure availability and reduce downtime. That way, the app can easily scale horizontally based on the number of users or when new features are added. Moreover, we can add microservices to run background operations continuously. In \texttt{baselayer}, this is used to run the application of course, but also the database migration manager, the webpack builder, websocket server, cron jobs, external logging, and \texttt{nginx}. In \texttt{SkyPortal}, this has been proven to be extremely useful when adding computationally expensive or long-running features such as the ingestion of GCN events with low latency, processing of observation plans, sending notifications and reminders, and programming recurrent API calls. These services will be discussed in further detail later in this paper.

\subsubsection{HEALPix-Alchemy}
\label{sec:hpa}

\begin{figure*}[t]
    \centering
    \includegraphics[width=6.5in]{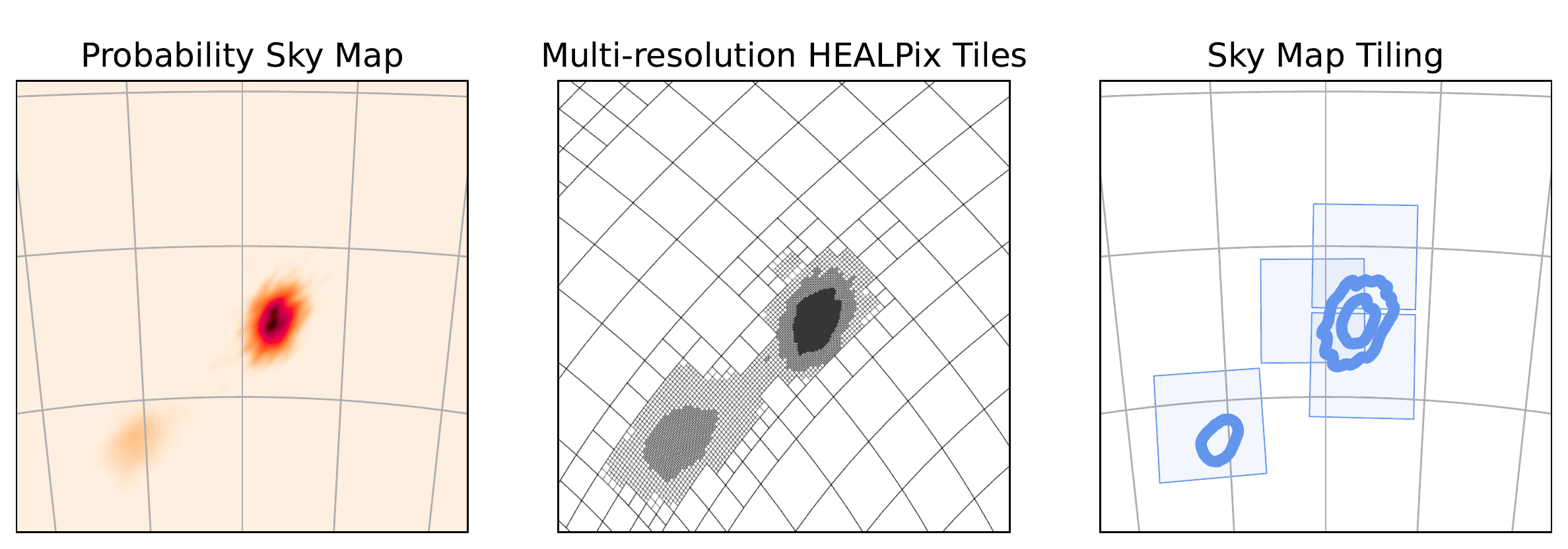}
    \caption{An example of the multi-resolution HEALPix sampling scheme for tiling localizations. The left panel shows an example heat map image representing the localization probabilities; darker, deeper colors represent higher probability density. The middle panel shows the boundaries of the multi-resolution HEALPix tiles on which the sky map was sampled. The right panel shows potential telescope field of views covering that skymap with the skymap confidence level contours included.}
    \label{fig:healpix-order}
\end{figure*}

To ensure calculations are sufficiently computationally expedient in \texttt{SkyPortal}, we represent the complex geometry of regions of interest (such as multi-messenger skymaps) and the instruments' fields of view using the Hierarchical Equal-Area isoLatitude Pixelization (HEALPix; \citealt{GoHi2005}) framework; to this end, we have developed a PostgreSQL extension enabling HEALPix-based cross-matches to make rapid processing on these large alert databases possible \citep{SiPa2022}, named \texttt{Healpix-Alchemy}. We briefly describe the innovation and its use within \texttt{SkyPortal} here, and encourage readers of this paper to see \citealt{SiPa2022} for further details.

In version 14, PostgreSQL introduced a new \emph{multirange} type, consisting of an array of ranges, with a fast aggregation function, which takes ranges as its input and returns their union as a multirange; when combined with the multi-order-coverage (MOC) representation of sky maps \citep{FeBo2014}, this tandem yields highly efficient database-side processing of skymap overlaps. 

Most multi-messenger events do not provide us with a precise localization to observe, but instead a map of probability density (see e.g., Fig.\,\ref{fig:healpix-order}) using MOC to divide the sky into tiles of diverse resolutions. Each tile has an associated probability density that the event occurred in it. MOC uses the HEALPix tessellation algorithm to divide the spherical observable sky into 12 diamond shapes of equal area. Each of them can be recursively divided into 4 to obtain the level of resolution needed to represent the variation of probability from one point of the sky to another; recursively dividing these tiles into 4 sub-diamonds increases the order and therefore the resolution.

At order 0, tiles have indexes from 1 to 12, and at order 1, from 1 to 64. For instance, tile 1 at order 0 is now made of tiles 1 to 4 at order 1, and tiles 1 to 16 at order 2, and so on recursively until the highest order is order 29, where a tile side size is only 0.38 milli-arcseconds; this minimum resolution is set by requiring that the corresponding pixel index can be stored as a 64 bit signed integer without overflow. 
Thanks to the recursive indexing schemes of the HEALPix algorithm, when saving MOC maps in the database, we can represent each tile as the range of tiles that it is composed of at order 29. This is convenient as it allows us to represent tiles as a probability density associated with a range set. To do so, \texttt{Healpix-Alchemy} combines the power of Healpix and SQLAlchemy, a python package used to communicate with PostgreSQL databases using python. \texttt{Healpix-Alchemy} uses the range sets feature of PostgreSQL databases to represent tiles and single points with 64 bits integers and provides the tools necessary to cross-match tiles and points. For software stacks that rely on relational databases, this range sets representation of HEALPix tiles is the best (and only viable) implementation to perform rapid spatial queries.

As MOC has become the standard representation for skymaps in multi-messenger astronomy \citep{fernique2022moc}, \texttt{Healpix-Alchemy} enables efficient queries of instrument footprints for rapid cross-matches with historical $\gamma$-ray burst, neutrino, and gravitational-wave skymaps.
Similarly, based on the coordinates of any object, observation, or galaxy, \texttt{SkyPortal} calculates and stores its associated HEALPix index at order 29 (which is the highest order, as stated above). This means that the spatial cross-match between sources, galaxies, and multi-messenger events is performed simply by looping over the localization tiles of an event localization to perform trivial intersection operations. Furthermore, thanks to the range set representation of PostgreSQL, performing the intersection and union of the HEALPix tiles representing regions of space is equivalent to a simple merging of sorted lists of integers \citep{singer2022healpix}. These intersection operations indicate if a given HEALPix index (a point) is contained in any of the range sets of the indexes corresponding to HEALPix tiles (a spatial region).

Using the same mechanism as the sky maps, \texttt{SkyPortal} also represents instrument fields using MOC representations of their footprint on the sky. The same basic union operations can be performed to determine which portions of a sky map (or in principle, objects and galaxies) an instrument is able to observe. Because the HEALPix algorithm has limits in resolution, the usage of MOC to define circles, polygons, and other objects with a basic geometry is in some sense an approximation \citep{fernique2015moc}, nevertheless, the MOC standard is an efficient way of representing instrument fields in a compact and generic manner. 

\subsection{Point Source Features}

While the line between features designed to interact with individual point sources and those designed to describe and interact with larger areas of the sky is blurry, we still find it useful as a broad diagnostic for understanding subsets of the features. Here, we describe the workflow for studying individual sources, from filtering to photometry to follow-up.

\subsubsection{Filters and Scanning}
\label{sec:filters}
With a full night of observing, ZTF sends $\sim$ 600,000--1,200,000 alerts per night, with over 2,000,000 alerts sometimes possible \citep{Patterson2018}; for comparison, Rubin Observatory expects to send $\sim$ 10,000,000 alerts per night. Given that the overwhelming majority of these alerts are not of interest to the average user, a combination of automated filtering and human scanning is used to reduce the alert list to a short list of sources deemed worthy of further attention.

Within \texttt{SkyPortal}, every science objective has a dedicated ``group'' of users, and one associated filter per group. These filters are defined within the brokers described in Section~\ref{sec:brokers}. 
As brokers are filtering the alert stream automatically in real-time, sources producing alerts that pass a given filter are known as \emph{candidates}. \texttt{SkyPortal} can be configured to immediately save all candidates that pass the filter to a group (a collection of sources), but in most cases, users will prefer to manually vet the stream to separate sources of genuine interest from false positives in a process referred to as candidate scanning.
All candidates passing a filter during a chosen time period (e.g., the past 24 hours or the past 3 nights) can be displayed rapidly on the Candidates page.
This page displays image cutouts for each candidate location from ZTF and other surveys, along with light curves and some additional information (coordinates, TNS cross-matches, etc.) and links to other resources.  
The user can peruse this list of candidates and save only those of interest based on this additional information.
Candidates that are not saved will reappear if the source continues to produce alerts that pass the filter in the future, but the user can also choose to hide any sources that are clearly false positives from their future scanning efforts.

\subsubsection{Photometry, Spectroscopy, and Photometric Series}

In addition to ingestion of photometry from alert streams provided by brokers, \texttt{SkyPortal} benefits from both public photometry services from Pan-STARRS Data Release II \citep{ChMa2016,Flewelling2018} and forced photometry services of both ZTF \citep{Masci2019} and ATLAS \citep{ToDe2018,SmSm2020}. These are available for querying through their API services. The visualization interface for photometry has been described previously \citep{WaCr2019}, but briefly, it is a \texttt{Bokeh} \citep{bokeh} implementation that allows one to view either magnitude or flux as a function of time, with tooltips providing basic information about the individual photometry points. There is also a panel that allows the user to specify a potential period to phase-fold the photometry, in case of periodic sources.

Statistics on each source are collected and updated upon the insertion of new data points.  This includes, e.g., the brightest magnitude, the time when the source was last detected,  and so on. These statistics are updated on insertion to avoid the costly re-calculation over hundreds of data points per source.  Such statistics are used to quickly filter sources based on their light curves, without needing to calculate each statistic for each source in each query. 

Another way to store large photometric datasets has recently been added to \texttt{SkyPortal}: the photometric series. 
This data product will be used by surveys with very high cadences and continuous coverage of the same field,
e.g., the Transiting Exoplanet Survey Satellite (TESS; \citealt{RickerTESS2014}) 
takes images at a maximum cadence of 20\,s over 27-day series. 
In this case, saving each of the $\sim 10^5$ data points as individual 
rows in the database becomes unmanageable, both in terms of storage
and in terms of communicating the data in and out of \texttt{SkyPortal}. 
Instead, we store the underlying data as HDF5 files and 
keep a queryable record in the main database of the file (along with some statistics). 

Spectra are uploaded either manually or via API. The spectra are associated with each source but the platform also enables the users to provide comments, allowing users to discuss 
either the source or the individual spectra. There are also standard line lists available to perform line identification, enabling classification.

\subsubsection{Finding Charts and Starlists}

To aid the rapid followup of sources of interest, \texttt{SkyPortal} enables photometric and spectroscopic finding charts to be generated on the fly. In interactive mode, a user selects a catalog image from DESI\,DR8, ZRF Reference Images, DSS2, or PS1, image size, and number of spectroscopic offset stars. Using a flux-weighted median position from all photometric detections, up to 4 nearby isolated stars from Gaia DR3 are selected as spectroscopic offset stars. The offsets are calculated using the parallax and proper motion of the stars from Gaia DR3 assuming the current datetime as the observing epoch. These offset stars can be copied from the frontend in different starlist formats. The finding charts can be generated programmatically and from the frontend (single click) with default parameters. A downloadable PDF is assembled and cached to facilitate fast observing run organization. 

\subsubsection{Periodograms}

For variable stars, \texttt{SkyPortal} enables an interactive periodogram analysis. An adapted javascript version of the Generalized Lomb-Scargle implementation \citep{2009A&A...496..577Z} allows for client-side computation of the periodogram. A user can change photometry filters and interactively select trial periods as well as visualize the folded light curve on the selection period and period aliases.

\subsubsection{Source Classifications}

\texttt{SkyPortal} offers a classification system that lets users tag sources with relevant labels. The namespace for classifications is set by taxonomies, which are collections of labels connected with each other using nested JSON format. \texttt{SkyPortal} sets a default Sitewide Taxonomy for use in all groups on the application based on the \texttt{tdtax}\footnote{\url{https://github.com/profjsb/timedomain-taxonomy}} repository.
Users can also post custom taxonomies (that adhere to a strict taxonomy schema) to specific groups, limiting their classifications' visibility to members of those groups.
Each classification has an associated probability that quantifies confidence in the label.

Based on the system built within the ZTF Variable star Marshal \citep{RoDu2021, CoBu2021}, \texttt{SkyPortal} allows users to post classifications from a drop-down menu one by one or use a slider interface that facilitates faster labeling. The slider interface initially shows the top-level classifications of the selected taxonomy. Each classification is accompanied by a slider to adjust its probability.
If the slider of a classification is moved to a non-zero probability, the child classifications of the parent will be shown along with additional sliders. This interface allows multiple classifications to be posted at once.

\texttt{SkyPortal} also features a classification voting system that facilitates collaborative labeling of sources in a prompt but informative way. Users can express their confidence in each existing classification visible to them by casting a thumbs ``up'' or ``down'' vote. 
When a user adds or deletes a source's classification or casts a vote, \texttt{SkyPortal} marks the source as ``Labeled.'' The user can later choose to view only unlabelled sources in a group to allow quick resumption of the labeling process.

\subsubsection{Follow-up}

\begin{figure}[t]
    \centering
    \includegraphics[width=3.5in]{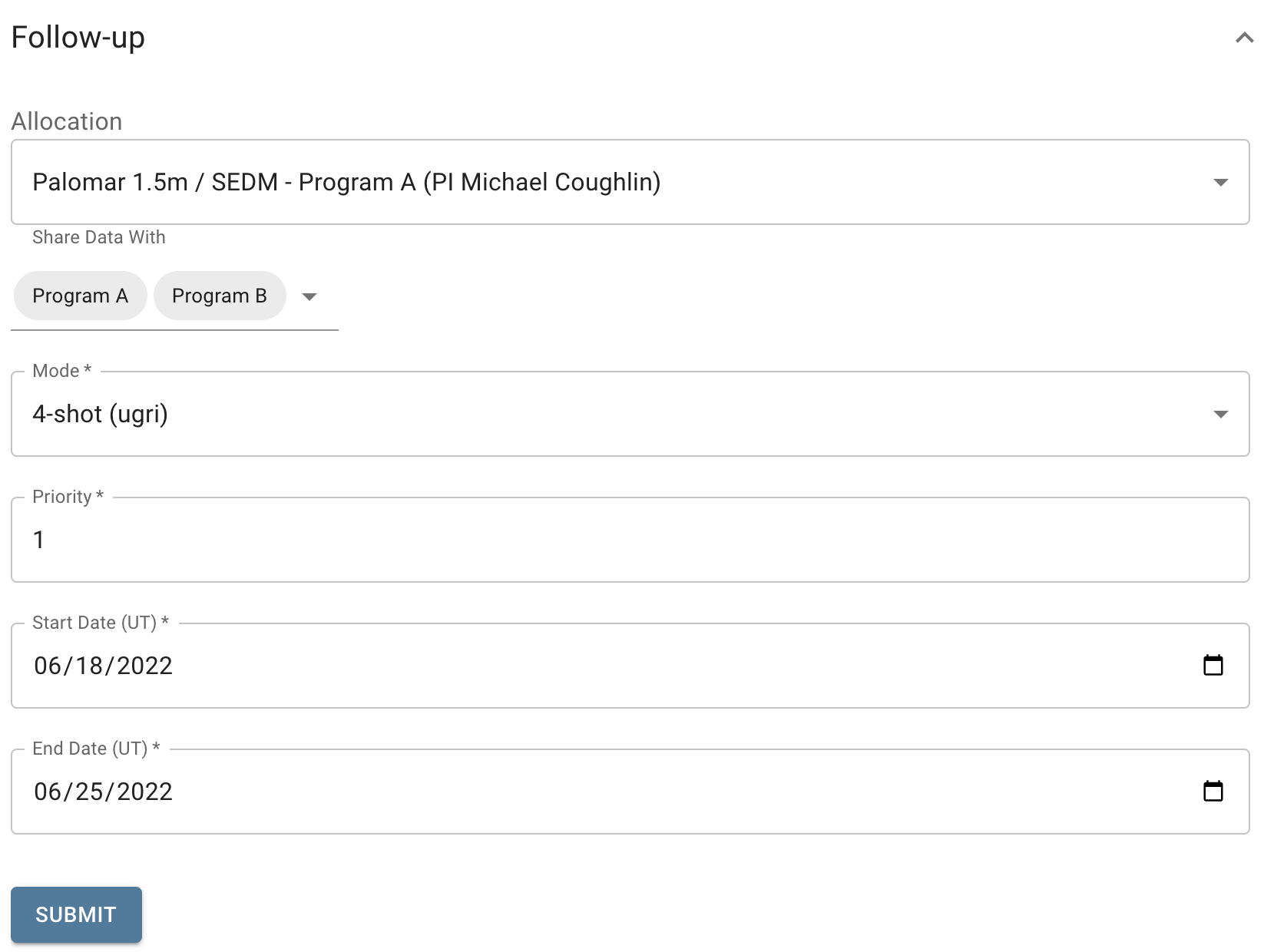}
    \caption{JSON-based follow-up interface.}
    \label{fig:follow-up}
\end{figure}

From within \texttt{SkyPortal}, API-based triggering is available for the telescope of the Las Cumbres Observatory network \citep{StBo2018}, the Katzman Automatic Imaging Telescope (KAIT; \citealt{LiFi2003}), the Liverpool Telescope (LT; \citealt{StSm2004}), Swift \citep{GeCh2004}, the SED Machine (SEDM;  \citealt{nblago18,RiNe2019}) on the Palomar 60 inch (P60) telescope, and the wide-field infrared transient explorer (WINTER; \citealt{LoBa2020}). Adding a new telescope follow-up capability is straightforward. Figure~\ref{fig:follow-up} shows an example drop-down menu customized for SEDM.

Groups that scan for candidates within \texttt{SkyPortal} often use the same platform to trigger photometric and spectroscopic follow-up in the same portal. Priorities, time-windows, and observational set-ups can be selected and edited. Once the data has been obtained, some systems such as the robotic SED Machine \citep{nblago18}, automatically reduce \citep{RiNe2019} and upload the spectra for visualization; as a deployment specific example, in the case of \texttt{fritz}, if a source spectrum is automatically matched to a thermonuclear spectrum (SNIa; \citealt{FrHa2021}), those spectra and classifications are also directly uploaded to the TNS as part of the Bright Transient Survey; \citealt{FrMi2019, Perley2020}. For other sources, human interaction allows for the classification (or rescheduling) of other targets, as well as the inclusion of detailed comments and discussions to share results and collaborate on particular objects.

\subsection{Non-point Source and Multi-messenger Features}

While many of the features described above have applicability to sources of all types, including infrastructure for follow-up, some of the features have been developed focusing on interactions with larger areas of the sky, as appropriate for many multi-messenger science cases. Here, we describe those features, focusing on their applicability for supporting multi-messenger source analysis and follow-up.

\subsubsection{GCN Events}

Much of the existing infrastructure for communicating multi-messenger observations originated from the GRB detection community in the 1990s \citep{bacodine}, with the decades-old ``Gamma-Ray Coordinates Network'' (GCN) continuing to play a central role in modern time-domain astronomy. This system includes the machine-readable ``GCN Notices,'' which uses three legacy communication formats (text, binary, and VOEvents; \citealt{allan2017ivoa}), each with its own communication protocols.
Recently, there have been concerted efforts to reimagine this system as a modern ``General Coordinates Network,'' transitioning to a unified communication protocol using open-source Apache Kafka, with the ultimate aim of relaying all data via the industry-standard JSON format\footnote{\url{https://gcn.nasa.gov/}}.

\texttt{SkyPortal} is fully integrated with this new GCN Kafka format and can ingest notices shared via this protocol. Which of the available streams should be ingested is specified in the configuration of the application. These GCN notices typically contain either full MOC sky maps \citep{fernique2022moc}, coordinates in right ascension and declination along with an associated error radius, or the vertices of a region shaped like a polygon. In all cases, a new entry is added to the database containing the localization itself (transformed to a MOC for efficient and standardized queries) and other available event properties such as localization area, median distance, or astrophysical probability. These events then can be subsequently queried, cross-matched to sources or galaxies, and used to coordinate subsequent follow-up. 

\begin{figure*}[t]
    \centering
    \includegraphics[width=6.5in]{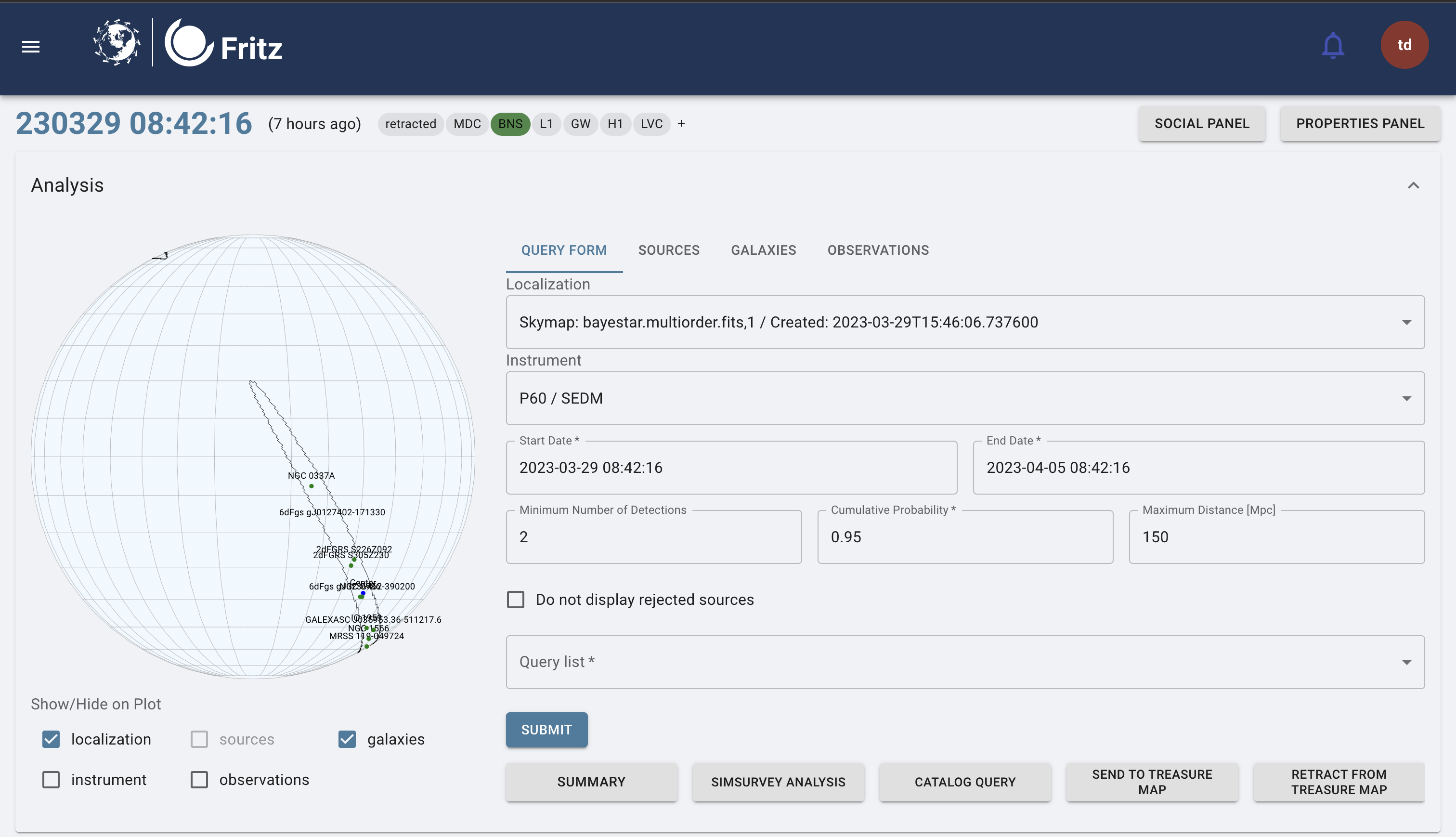}
    \caption{GCN Page - Analysis Section: here, users can visualize the localizations associated with an event, and query the sources, galaxies, instrument tiles (for a given allocation on an instrument), and observations contained in the different localizations, while restraining their searches based on several parameters: cumulative probability, number of detections, distance in Mpc, and first and last detection dates. Other features accessible in this section allow to query catalogs of candidates from different alert brokers, run simulations with \texttt{simsurvey}, and generate text-based summaries similar to GCN circulars with added version control. It is also possible to submit the executed observations to \texttt{treasuremap.space} \citep{WyTo2020}. Side panels can be opened that contain the properties of the event and localizations, lightcurves, GCN notices and circulars, as well as a comment section for users to communicate their results.}
    \label{fig:GCN_analysis}
\end{figure*}

\texttt{SkyPortal} supports the analysis of these events by allowing for different queries within the localization map: telescope observation footprints, galaxies, and transient sources. The analysis page of each event can display the error regions in a spherical projection, and can additionally display the results of the queries (see an example in Fig.\,\ref{fig:GCN_analysis}). Moreover, the sources returned in the query will appear listed in the \textit{Sources} table of the event page, enabling further visual inspection. The live interaction that occurs in the \textit{Sources} menu has shown to be essential to enable candidate vetting and follow-up prioritization. For example, by clicking the arrow next to the name of the source or candidate, \texttt{SkyPortal} displays the available photometry, as well as cutouts of the coordinates of the source. Once a candidate association has been confirmed or ruled out, its status can be updated and comments can be left for future reference or internal communication amongst collaborators. Finally, \texttt{SkyPortal} can use this information to generate automated GCN circulars which summarize observations performed by telescopes that overlap the event, the associated sources detected by those observations, and the estimated event localization probability covered by those observations.

\subsubsection{Observation: Planning and Execution}

A central component of multi-messenger astronomy is the efficient coordination of follow-up for a given event, especially for gravitational-wave and GRB triggers which are initially poorly localized. Astronomers may have access to both wide field-of-view instruments and narrow field-of-view instruments of different depth and wavelength coverage. Developing strategies for optimal follow-up with a particular telescope has been an area of active research in recent years \citep[e.g.,][]{CoTo2018,CoAn2019, AlCo2020}.

\texttt{SkyPortal} has extensive functionality to automatically perform these optimizations in a user-friendly GUI. Upon ingestion of localizations, \texttt{SkyPortal} can generate optimized observing schedules using the open-source ``Gravitational-wave Electromagnetic Optimization'' (\texttt{gwemopt}) software package \citep{CoTo2018,CoAn2019}. \texttt{gwemopt} can generate custom plans for individual telescopes, or broader network plans which coordinate multiple telescopes simultaneously. It can either generate an agnostic sky-tiling schedule, or use internal catalogs to generate a galaxy-targeting schedule while accounting for user-tunable factors such as airmass, start and end time, or available filters. 

Plans are typically generated by default at the time of event ingestion, with the option for further plan generation via a configurable web form. Each of these plans can be visualized on the event page, overlaid with the initial localization. Observation plans can be triggered automatically, but can also be triggered manually from the GCN Event page. 

To assess the efficacy of the generated plans, we use the open-source simulator software \texttt{simsurvey}\footnote{\url{https://github.com/ZwickyTransientFacility/simsurvey}} \citep{FeNo2019} (and are working with the author of \texttt{skysurvey} to integrate it\footnote{\url{https://github.com/MickaelRigault/skysurvey}}). In particular, we use {\tt simsurvey} to simulate transients of various types (kilonovae, GRB afterglows, supernovae, etc.) to estimate the probability of detecting one of these transient types within the skymap, given the proposed observation plan. For kilonovae, the optical/NIR counterparts to binary neutron star mergers generated from the radioactive decay or $r$-process elements \citep{Me2017}, we use a \texttt{POSSIS}-based \citep{Bul2019} grid of kilonova models spanning the plausible parameter space for kilonovae from binary neutron star and neutron star--black hole mergers \citep{DiCo2020, Bul2023}. For GRB afterglows, we use \textit{afterglowpy} \citep{RyEe2020}, an open-source computational tool modeling forward shock synchrotron emission from relativistic blast waves as a function of jet structure and viewing angle. This analysis accounts for properties of the skymaps (i.e., probability in right ascension and declination, distance in the case of gravitational-wave skymaps), as well as properties of the observations (i.e., right ascension and declination for each field, observation time, limiting magnitude, filters, etc.). 

\begin{figure*}[t]
    \centering
    \includegraphics[width=6.5in]{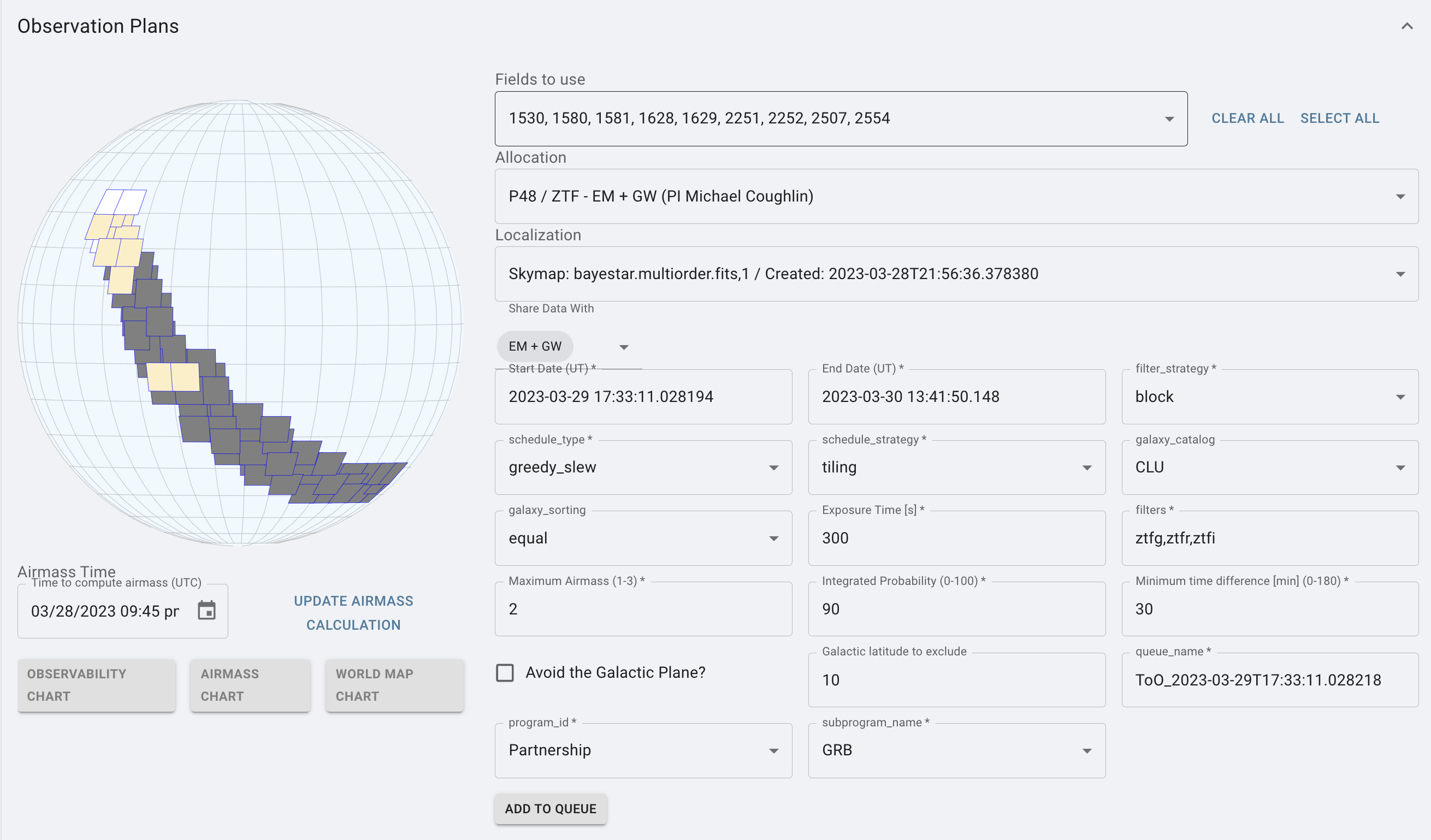}
    \caption{GCN Page - Observation planning: here, users select an allocation, the fields of the instrument of the allocation that overlap with a localization, as well as pick the parameters to use from the MMA API class of the instrument. Multiple observation plans can be created simultaneously and combined if necessary. Moreover, buttons allow users to download observability and airmass charts, as well as recompute airmass in real time to show which tiles of the skymap are observable. It is also possible to submit the proposed observations to \texttt{treasuremap.space} \citep{WyTo2020}. Once created, all observation plans will be listed by instrument, where they can be updated, sent, or deleted to/from an instrument queue.}
    \label{fig:GCN_obsplan_form}
\end{figure*}

To send back data from the observations based on these plans, telescope teams can use  the API of\texttt{SkyPortal}, or add them directly on the Observation Page. Observation plans will automatically be cross-matched with the newly added observations, so they can appear on the list of observation plans for the event. If all observations scheduled in a plan are finished, the observations will be labeled as completed.
Due to the elusive (red and fast-fading) nature of gravitational wave and GRB counterparts, several consecutive follow-up campaigns could result in a limited number of counterpart detections. Thus, one of the main science cases in multi-messenger astronomy is using upper limits to constrain transient properties \citep{CoDi2019b,CoDi2020}.
To quantify our sensitivity to electromagnetic counterparts from our past observations, we again use \texttt{simsurvey}. We inject light curves consistent with the distance and location information contained in the skymaps and calculate the detection efficiency (i.e., the ratio between the number of detected transients and the number of injected transients) based on the field-by-field limiting magnitudes of our uploaded observations. We can then report these detection efficiencies, along with the median depth of our observations promptly via GCN.

\subsubsection{Galaxy Catalogs}

Due to the relatively local nature of multi-messenger astronomy, both follow-up focusing on galaxy catalogs \citep{ArHo2017,VaSa2017} and comparison of transient locations to catalogs is important. 
Within the database, galaxies can have several attributes: name, alternative name, catalog name, RA, Declination, redshift ($z$), and distance in Mpc (if known), as well as other properties and their associated margin of error if provided. Concerning the usage of catalogs, as deployment specific examples, in \texttt{fritz}, we use the Census of the Local Universe (CLU) \citep{CoKa2017} catalog, which is complete to 85\% in star formation and 70\% in stellar mass at 200\,Mpc; this catalog currently remains proprietary. In \texttt{icare}, GLADE+ \citep{dalya2021vizier}, an open-source catalog containing more than 22 million galaxies, complete up to 44\,Mpc and nearly 90\% complete at even 500Mpc, is used. The latter is a combination of six separate astronomical catalogs: GWGC, 2MPZ, 2MASS XSC, HyperLEDA, and WISExSCOSPZ. 

While we focus on these two catalogs, the API interface makes ingestion of other galaxy catalogs (and therefore cross-matching) straightforward, with a minimal requirement of right ascension, declination, and galaxy name. As one can imagine, the large size of these catalogs drives the need to perform efficient cross-matches between objects, multi-messenger event localizations, and these galaxies; our innovations with \texttt{Healpix-Alchemy} has helped, however, these are challenges that we are still working on to date, with performance improvements in the roadmap for O4. However, initial versions suitable for production performance are already implemented and functional.

\subsection{Interpretation and Coordination Features}

There are some features in the application with utility irrespective of the source(s) of interest. These include person-power coordination, notification, and analyses. Here we describe the \texttt{SkyPortal} infrastructure built to address such components of the MMA workflow.

\subsubsection{Shift Scheduling}

The scheduling of person-power is essential to implementing a robust multi-messenger or time-domain astronomy campaign; time is of the essence when it comes to triggering the follow-up of a source or planning the observations necessary to cover the localization of an event. Within \texttt{SkyPortal}, we have introduced the concept of ``shifts,'' whereby each day is divided into shorter periods---usually in 4 periods of 6 hours---where teams of follow-up advocates and ``shifters,'' whose time zone best accommodates such a period, are assigned. 

\begin{figure}[b]
    \centering
    \includegraphics[width=3.5in]{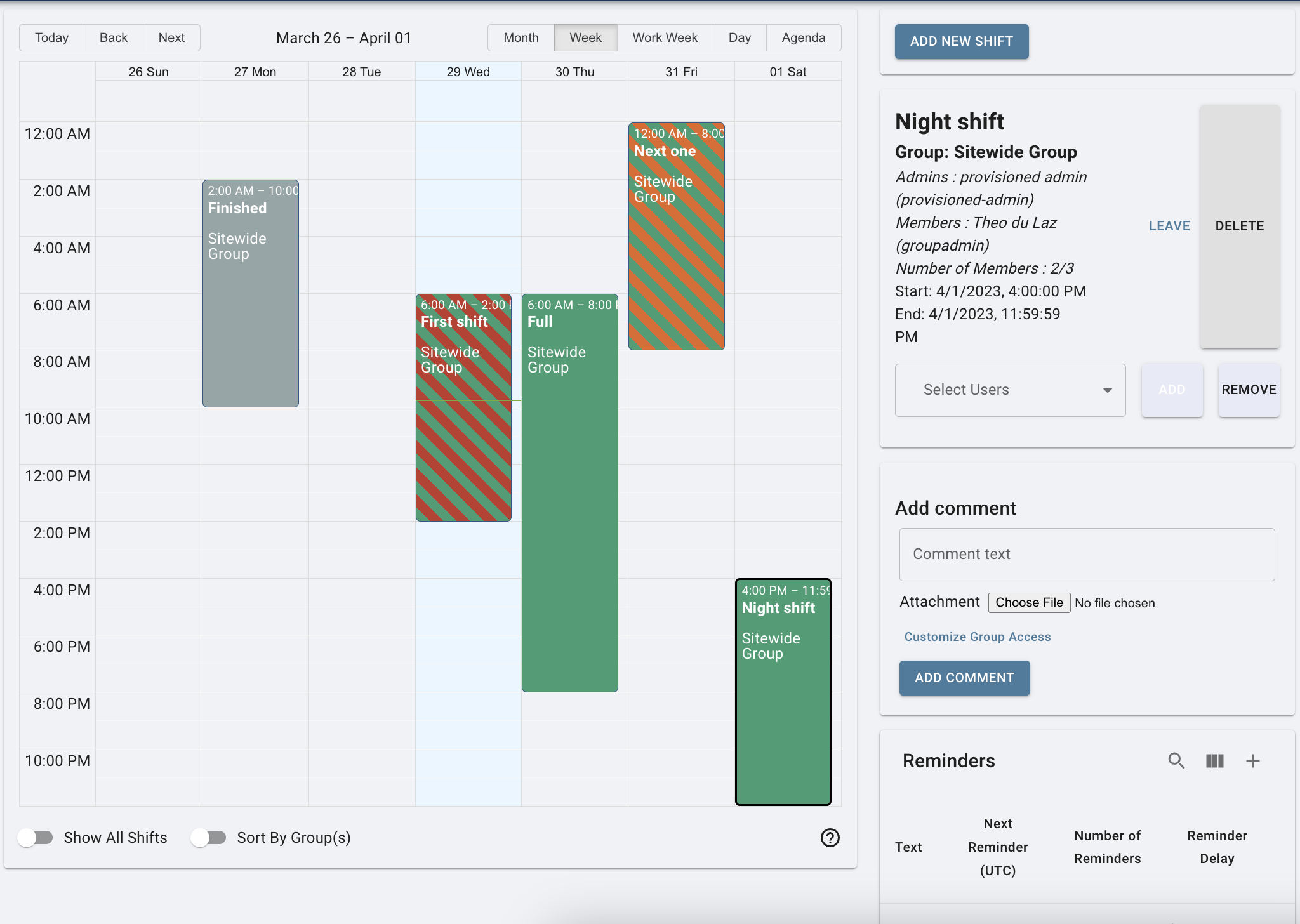}
    \caption{Shifts Page: users can see the shifts on a calendar and click on them to display more information and interact with them, as well as create new shifts.}
    \label{fig:shifts}
\end{figure}

A page dedicated to the shifts has been added to \texttt{SkyPortal} to allow the creation, visualization, and management of shifts through a Google calendar-like page. On the shift calendar, one can: create a shift, join or leave a shift, delete a shift, comment on a shift, and add or remove people (for those who have the admin role). Each shift has an associated group and max shifter capacity. If said capacity is not reached within 48\,hr before the start of the shift, this will be indicated on the calendar, and members of the group will be notified as well. Moreover, our experience during the previous O3 campaign showed us that the planning of a shifter could change at the last minute, preventing them from covering their shift and breaking the continuous monitoring of new events. To solve this problem, a feature allowing a shifter to request a replacement from other members of their group by notifying them was added. When selecting a shift on the calendar, one can see the shift management panel, the comments panel, the reminders panel, as well as the shift summary panel at the bottom of the page. On the shift summary element, a list of events that occurred during the selected shift---if any---will appear. By clicking on an event, one can see the list of sources contained within the latest localization of the event. The sources listed are the ones that have been first and last observed within one week after the first detection of the event. When leaving a comment associated with a shift in its comment section, one can mention/tag events, sources, and other users. This feature allows us to centralize and gather the work of shifters on \texttt{SkyPortal} in real-time while associating their messages with objects and events. This capability has replaced some need for collaborations to rely on external tools such as Slack, which has been the option of choice so far for this type of scheduling process.

\subsubsection{Notification framework}

As previously mentioned, we rely on notifications to keep users informed about recent activities within the app in a user-tunable fashion, given that not all users need to be notified of the same activities. To accomplish user-defined notifications, we built a notification framework on \texttt{SkyPortal}, where each user can specify their needs on the user profile page. On this page, one can choose to be notified of the following:

\begin{itemize}
\item When a given classification is added to any source, based on a list of classifications the user is interested in.
\item When sources tagged as ``favorites'' record different activities selected by the user, such as new comments, new spectra, and new classifications.
\item When a new GCN event or new GCN notice for an event already in the platform is ingested. The user can select which GCN notice types they want to be notified for, as well as specify properties of either the event itself (e.g., classification as a binary neutron star) or the skymap (e.g., size of the localization region).
\item When a new facility transaction is added, including follow-up requests and observation plans.
\item When a user is mentioned by another user in a comment.
\end{itemize}

\begin{figure}[t]
    \centering
    \includegraphics[width=3.5in]{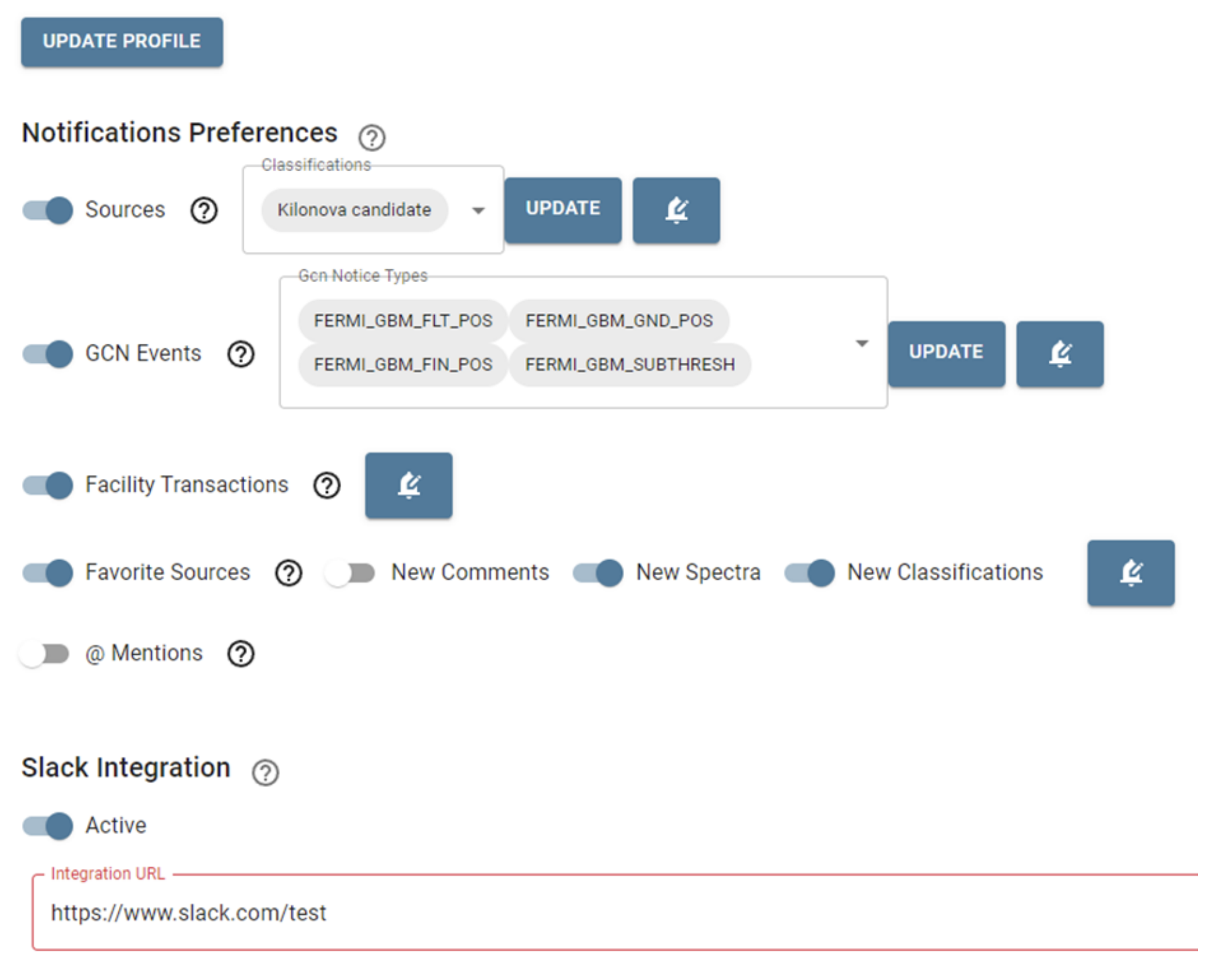}
    \caption{User profile - Notification preferences: users can activate or deactivate notifications, specify parameters on which to notify, as well as choose where to be notified by clicking the bell icon next to each notification type.}
    \label{fig:notifications}
\end{figure}

For each of these notification types, in addition to the in-app notifications, the user can choose to be notified by email, Slack, SMS, WhatsApp, and/or phone call. By email and Slack, a user will always be notified. When the SMS, Whatsapp, and/or phone call option is selected, users must choose if they want to be notified while on shift, and/or every day but during a specific time period (i.e., from 8 AM to 10 PM local time). For instance, during a recent observation campaign, shifters activated notifications for new events of the notice types that the campaign was interested in (in that case, Swift notifications specifically). Moreover, the possibility of ``bot'' notifications to create alert channels on Slack was added; these work through the user providing a slack channel URL to interact with as a webhook. We also automatically subscribe some users to receive certain critical notifications; for instance, \texttt{SkyPortal} always notifies the admins of an allocation when a new facility transaction has been created on the instrument they are admins for. This is done to ensure that new follow-up requests or observation plan submissions cannot be missed by those responsible to trigger the associated observations on their telescopes.

\subsubsection{Analysis Platform}
\label{sec:analysis-service}

A common way to characterize transients in optical astronomy is through the analysis of photometric, multi-band light curves. A variety of Bayesian inference frameworks have sprung up to meet the needs of transient characterization, including MOSFiT \citep{GuNi2018}, the ``Nuclear Physics - Multi-messenger Astronomy'' (NMMA) framework \citep{PaDi2022}, amongst others. These frameworks are used to calculate Bayes factors for the purpose of model comparison as well as posteriors for transient characterization. While time-of-discovery analyses can reside within the purview of brokers, analysis of follow-up photometry and spectra can provide critical insights.

\begin{figure}[bh]
    \centering
    \includegraphics[width=3.3in]{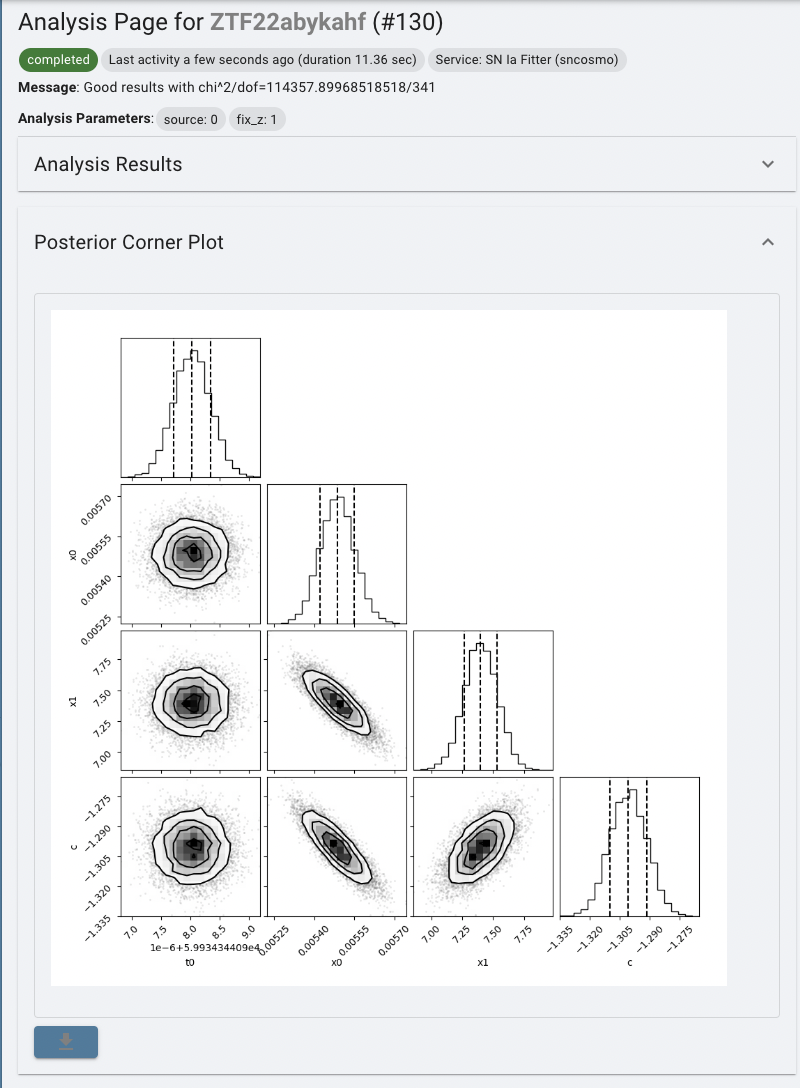}
    \caption{Analysis posterior for a fit to ZTF\,22abykahf with the Type Ia supernova analysis service.}
    \label{fig:analysis}
\end{figure}

\texttt{SkyPortal} now enables a wide range of analyses on individual sources, viewing analysis packages as standalone web services capable of receiving source data and returning analysis summaries (in JSON format), parameter posteriors (in \ttt{xarray} \ttt{arviz.InferenceData} format [\citealt{Kumar2019}]; Fig.\,\ref{fig:analysis}), and/or diagnostic plots. Site administrators first create analysis services, by providing a URL (and authentication credentials) to an existing service, the schema of required and optional parameters, and data types to be sent from \texttt{SkyPortal} to the analysis service. When an analysis is initiated (either via the front end or via API), all data viewable by the user in the analysis data types are packaged and sent to the analysis service. Via a unique unauthenticated webhook, the analysis service then asynchronously returns analysis results. (Analysis results generated without \texttt{SkyPortal}-packaged source data and without a webhook can also be saved by an authenticated token.)
The results are stored in a persistent datastore by \texttt{SkyPortal} and can be retrieved by analysis id or by querying for analysis results by source name. Posterior corner plots and other returned plots are also rendered on individual analysis pages.

While the analysis platform enables remote 3rd party services, \texttt{SkyPortal} also ships with pre-built analysis services that are run locally as microservices. A supernova service wraps the \texttt{sncosmo} \citep{BaBa2016} fitter and an afterglow/kilonova service wraps the \texttt{nmma} \citep{PaDi2022} fitter.

\subsubsection{AI Summarization \& Embeddings}

\begin{figure}[t]
    \centering
    \includegraphics[width=3.3in]{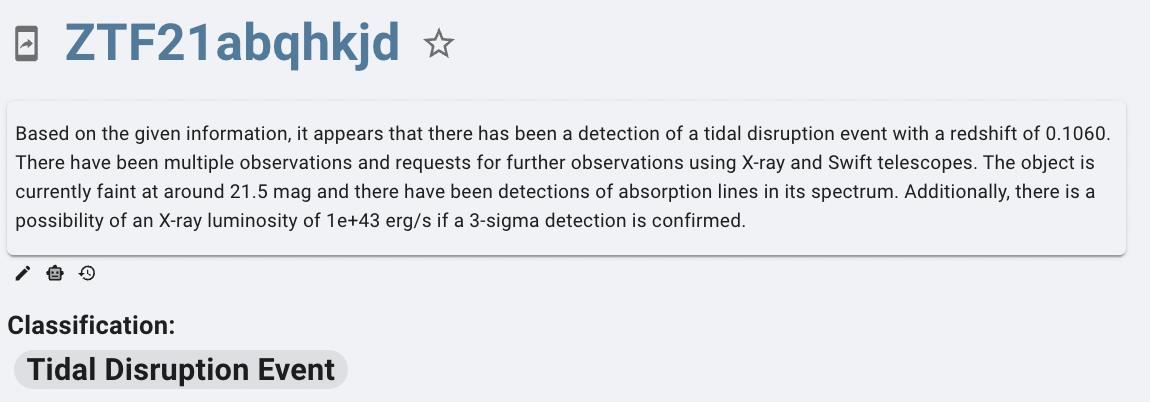}
    \caption{Source page visualization of the AI-generated summary of using the ChatGPT 3.5 completion service on ZTF\,21abqhkjd \citep{2021TNSAN.258....1Y}.}
    \label{fig:summary}
\end{figure}

One challenge for end-users in the face of hundreds (or thousands) of active events is maintaining a current understanding of the state of knowledge and activity for each event.  To facilitate ``quicklook'' insights into sources, following the emergence and general availability of highly capable large language models (LLMs), the \texttt{SkyPortal} team released a pre-built analysis service to provide human-readable summaries of individual sources, using redshift, classifications, and comments. When such an analysis is initiated on a source, the summarization service formats available source data along with a predefined prompt requesting a summary; the OpenAI API (\texttt{openai.ChatCompletion}) returns a summary which is then displayed on the source (fig.\,\ref{fig:summary}). This summary can be easily edited, with the summary history being saved. A sitewide API key to OpenAI can be used or individual users can configure their own key and specific completion parameters. \texttt{SkyPortal} defaults to use the \texttt{gpt-3.5-turbo} model but users can opt to employ the more capable (but currently slower) \texttt{gpt-4} model.

Once a summary is created, an OpenAI embeddings service (using the \texttt{text-embedding-ada-002} model) is called with the summary that returns a vector embedding of the source in $n=1536$ dimensions. These embeddings are saved in a \texttt{pinecone} database\footnote{\url{https://www.pinecone.io/}}. Using these embeddings, similar sources can be retrieved by rank-ordering cosine distance metrics or natural-language queries (first embedded, then compared against the summary corpus) can be answered. Future embeddings will incorporate photometry, spectra, and annotations to further facilitate AI-guided exploration of \texttt{SkyPortal} datasets.

\section{Deployments, Science Validation and First Results}
\label{sec:science}

In this section, we describe the specifics of two \texttt{SkyPortal} implementations, \texttt{fritz} and \texttt{icare}, as well as a worked example of a an ``offline'' search for $\gamma$-ray burst afterglows from Fermi-GBM with ZTF. \texttt{SkyPortal} implementations mainly differ on (i) where they are deployed and (ii) the data stream that they receive, and so in these cases, the brokers that they rely upon to filter and post results from ZTF. For (i), \texttt{fritz} is deployed on Google Cloud Platform, while \texttt{icare} runs on dedicated hardware on site at IJCLab in Orsay, France. For (ii), we describe these in the following subsection.

\subsection{Alert Brokering}
\label{sec:brokers}

Several \texttt{SkyPortal} instances rely on alert brokers to populate their databases with candidates that users can scan and save as sources (or reject) for their groups. In the case of \texttt{fritz}, it uses \texttt{Kowalski}\footnote{\url{https://github.com/skyportal/Kowalski}}, an open-source, multi-survey data archive and alert broker \citep{DuMa2019}, for performing filtering and cross-matches, including rejection of bogus objects \citep{DuMa2019}; some science groups also rely on \texttt{AMPEL} \citep{Nordin:2019kxt}. In the case of \texttt{icare}, it relies on \ztfink\, \citep{AiAl2022}, including its filtering capabilities targeting fast transients, to select promising candidates.

\subsubsection{\texttt{Kowalski}}

\texttt{\texttt{Kowalski}} comprises a non-relational (NoSQL) \hbox{MongoDB}\footnote{\href{https://www.mongodb.com/}{https://www.mongodb.com/}} database to store ZTF alerts and other catalogs, and an API layer that allows external users (such as \texttt{fritz} itself, or the users using it) to interact with it. Among other benefits, MongoDB provides fast execution times ($\sim\log(N)$) for standard operations, allows for efficient positional queries, and saves individual entries in the binary JSON format---well-matched with the \texttt{AVRO} format of ZTF alerts. A dedicated \texttt{Kafka} consumer in \texttt{\texttt{Kowalski}} consumes the ZTF alert stream at IPAC in real-time and ingests alerts into the database, where they are stored as ``documents.'' \texttt{\texttt{Kowalski}} provides an infrastructure relying on parallel computing to filter and enhance many alerts at once prior to ingestion. Basic filters can be specified in the configuration, on top of which user-based filters can be defined for the different instruments handled by \texttt{Kowalski}. Those can be sent to \texttt{Kowalski} using its API, allowing the development of GUIs (like the dedicated filter page on \texttt{Skyportal}) where users can define filters in a human-readable JSON format (see Sec. \ref{sec:filters}). 

Users define filters in the Metaweb Query Language (MQL) syntax, so that they can be run on the alerts stored in \texttt{\texttt{Kowalski}} as ``documents.'' A filter takes the form as a MongoDB ``aggregation pipeline'' that is tuned to a specific scientific objective. An aggregation pipeline is a list of dictionaries, each representing a specific stage in the filtering process. A stage performs an operation on the document, such as calculating a value (e.g., the age or color of the transient) or filtering the document (e.g., only pass documents with color\,$>$1\,mag.). All documents are fed to the first stage, but only the output of a stage is fed to its subsequent stage---this ensures that only documents of interest are retained by the end allowing for efficient automated filtering of alerts. 

As part of the filtering document, users can specify which alert fields can be posted to \texttt{SkyPortal} as annotations, allowing for quick analysis of source properties. For alert enhancement, \texttt{Kowalski} can be configured to cross-match alerts with other catalogs and candidates from the other instruments or run machine learning inference on them. The user-defined filters are applied after the cross-matches and machine learning scores are added, meaning that filtering can be done based on those values and not only the original alert fields. Based on the filtering results, an alert sentinel will post candidates and annotations. Once the enhanced alerts are ingested into \texttt{Kowalski}'s database, users can query them by running MongoDB Query Language (MQL)-based queries using \texttt{\texttt{Kowalski}}'s API. The package \texttt{penquins}\footnote{\url{https://github.com/dmitryduev/penquins}} provides a python client to interact with the \texttt{\texttt{Kowalski}} API. \texttt{\texttt{Kowalski}} is containerized using \texttt{Docker} allowing for simple deployment on the cloud or on-premises. 
 
For \texttt{fritz}, \texttt{Kowalski} serves as an ``upstream" aggregation pipeline that combines all photometric history, cross-match data, and ML scores for a ZTF alert available into a single document is run and then passed to the user-defined filter. Filters saved on \texttt{fritz} are version controlled and can be viewed/updated on a dedicated tab.
 
The instance of \texttt{\texttt{Kowalski}} that serves as the backend for \texttt{fritz} is deployed on-premises at Caltech. As of March 2023, it stores $\sim$38 TB of data including \hbox{$\sim$555} million ZTF alerts and \hbox{$\sim$20} other catalogs (e.g., 2MASS, \citealt{Cutri2003book}; ALLWISE, \citealt{Cutri2012wise}). It processes millions of requests daily, received either from \texttt{fritz} or $\sim$10 individual users for whom direct access is allowed. It handles candidate filtering for $\sim$80 ZTF experiments (e.g., the Bright Transient Survey; \citealt{FrMi2019, Perley2020}) by running filters saved on \texttt{fritz} (see Sec. \ref{sec:filters}) on all new ZTF alerts in real-time. In addition to ZTF, this instance also manages the alert stream for Palomar Gattini-IR \citep{De2020pasp}---an infrared time domain survey---in a similar manner and may do the same for WINTER \citep{Lourie2020spie}. These figures are summarized in Table~\ref{table:numbers}.

\begin{table}[h]
\centering
\caption{\texttt{Kowalski}}
\begin{tabular}{|l|c|}
\hline
Data Stored & $\sim$38 TB \\
\hline
ZTF Alerts & $\sim 5.55 \times 10^{8}$ \\
\hline
Other Catalogs & $\sim 20$ \\
\hline
\end{tabular}\\
Key numbers for the \texttt{Fritz} instance of \texttt{Kowalski}.
\caption{\texttt{fritz}}
\begin{tabular}{|l|c|}
\hline
Photometry & $\sim$ 400 million  \\
\hline
Candidates & $\sim$ 7.3 million \\
\hline
Sources & $\sim$ 420,000 \\
\hline
Spectra & $\sim$ 11,000 \\
\hline
Groups & $\sim$ 500 \\
\hline
Users & $\sim$ 320 \\
\hline
Tokens & $\sim$ 160 \\
\hline
Filters & $\sim$ 80 \\
\hline
Telescopes & $\sim$ 70 \\
\hline
Instruments & $\sim$ 85 \\
\hline
Comments & $\sim$ 170,000 \\
\hline
Annotations & $\sim$ 3.2 million \\
\hline
Thumbnails & $\sim$ 12.2 million \\
\hline
GCN events & $\sim$ 3,100 \\
\hline
\end{tabular}\\
Key numbers for the \texttt{Fritz} instance of \texttt{SkyPortal}
\label{table:numbers}
\end{table}

\subsubsection{AMPEL}

In addition to \texttt{\texttt{Kowalski}}, some science working groups in ZTF use \texttt{AMPEL} \citep{Nordin:2019kxt} to process the alert stream and display the results on \texttt{fritz}.

\texttt{AMPEL} divides its processing into four tiers. Each tier is handled by a set of \textit{units} (pluggable, configurable modules) that each perform a specific task. At Tier 0 (\textit{ingest}), alert packets are filtered, and selected alerts are decomposed into constituent data points (e.g., individual photometric points) and stored. At Tier 1 (\textit{combine}), data points are combined into \textit{states} (e.g., light curves). At Tier 2 (\textit{augment}), data points and states are enhanced with auxiliary or derived information, e.g., catalog matches for photometric points or model fits for light curves. At Tier 3 (\textit{react}), documents are aggregated by \textit{stock} (transient object being tracked) and can be reported to an external service, e.g., to request follow-up for the N most interesting objects, or simply provide a target list for a traditional scanning workflow. Users can customize each of these tiers to create a \textit{channel}, a real-time analysis plan specific to their intended science program. This could consist of e.g., a custom alert filter, datapoint association scheme, set of light curve analyses, and reporting scheme. Tiers 0--2 are driven by the alert stream, while Tier 3 is run at fixed intervals.

\texttt{AMPEL} integrates with \texttt{SkyPortal} via \texttt{SkyPortalPublisher}, a Tier-3 unit distributed as part of the \texttt{Ampel-ZTF} package\footnote{\href{https://github.com/AmpelAstro/Ampel-ZTF}{https://github.com/AmpelAstro/Ampel-ZTF}}. Each stock that \texttt{SkyPortalPublisher} processes is saved as a candidate for a set of filters. It assumes a one-to-one mapping between \texttt{AMPEL} channels and (dummy) filters stored on \texttt{fritz} by default, but a custom set of filters may be configured instead. Additionally, a stock is saved as a source in a set of target groups if so configured. Light curve analysis results from Tier 2 are serialized to JSON and posted as comments on the source. This behavior predates the analysis platform described in Sec.~\ref{sec:analysis-service}, and we expect to migrate the analysis reporting of \texttt{AMPEL} to this service in the near future. If photometry and image cutouts were loaded as part of the Tier 3 process, these are posted as well. This makes it possible to use \texttt{AMPEL} with a standalone \texttt{SkyPortal} instance, but this is disabled in the ZTF instance, as competing backends tend to confuse users and lead to unnecessarily high database loads. Finally, \texttt{SkyPortalPublisher} keeps a record of its communication with \texttt{SkyPortal} in the journal associated with each stock, and only attempts to post if it was updated since the last successful attempt.

Some science programs require two-way communication, e.g., the ZTF Nuclear Transients group uses classifications that human astronomers post to \texttt{Fritz} in their \texttt{AMPEL}-based scanning workflow. \texttt{AMPEL} enables this via e.g., \texttt{FritzReport}, a Tier-3 complement stage that injects a source record from \texttt{SkyPortal} into stock records before they are provided to downstream units.

\subsubsection{Fink}

The GRANDMA collaboration uses the Fink broker \citep{MoPe2020} to select promising alerts from the public ZTF alert data stream and display the results on \texttt{icare}.

The current Fink platform works in four steps. First, alerts from multiple streams are continuously ingested and stored on disk. Second, alerts satisfying the quality cuts defined by the broker team are processed by a set of science modules. These science modules---currently a dozen, spanning solar system objects to galactic and extragalactic science---are independent processing units developed by the community of users, and deployed in the Fink platform. They can work on a single input alert stream, or combine several streams together. These science modules enrich the initial alert packets using several techniques such as cross-match with external catalogs of astronomical objects, or classification using machine or deep learning based algorithms. All added-values are made public. Third, enriched alert packets are filtered based on their content, and the most promising events are redistributed to the scientific community in real-time. The filtering is again community-driven, and users design and deploy filters to receive tailored information in real-time. Finally, all enriched alert packets are stored in a database for permanent access and for further analyses. Fink exposes a number of services to the users to design modules, filters, and access the data\footnote{\url{https://fink-broker.readthedocs.io}}.

Fink integrates with \texttt{icare} with the \texttt{SkyPortal Fink Client}\footnote{\url{https://skyportal-fink-client.readthedocs.io}}. Alerts are first polled using the Fink Livestream service, and then pushed to \texttt{icare} using core functionalities of \texttt{SkyPortal}. The \texttt{SkyPortal Fink Client} runs as an optional microservice inside \texttt{icare} which simplifies the deployment and the maintenance for users. We expect to migrate this optional microservice to the main \texttt{SkyPortal} codebase in the near future so that it can benefit the \texttt{SkyPortal} community at large.

In the context of GRANDMA, three specific Fink filters targeting alerts associated with fast transients have been deployed \citep{AiAl2022}. One of the filters is based on the output score of a dedicated science module implementing a fast transient classification algorithm, and all filters were used in two subsequent follow-up campaigns of optical transients to search for orphan kilonova \citep{Biswas:2022ocj}. Selected alerts are automatically published into \texttt{icare}, where continuous human scanning takes place (see below for scanning capabilities within the platform). Astronomers would then decide to trigger further observations based on the received alerts content, but also using external information collected by the GRANDMA collaboration and ingested into \texttt{icare} directly by other means.

\subsection{Example Science Results}
\label{sec:science}

\begin{figure*}[t]
    \centering
    \includegraphics[width=6.5in]{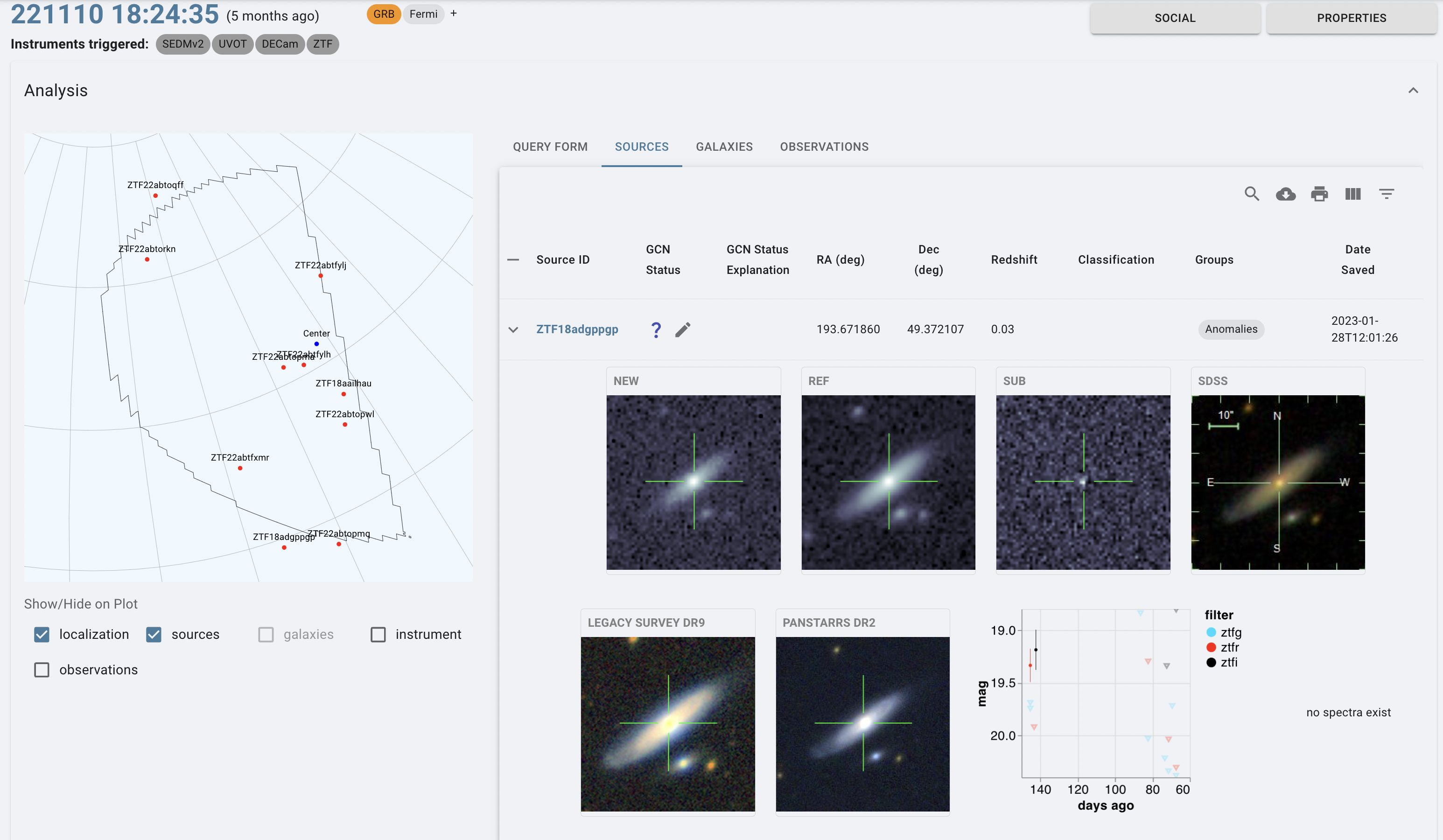}
    \caption{Analysis section from the GCN event page of GRB\,221110A. The GBM region is uploaded after correcting for Earth occultation, which produces a cut on the shape of the original map. The sources detected in the region covered by ZTF are shown as red dots, with their names. The \textit{Sources} menu allows for in-depth scanning of the candidates, showing cutouts of ZTF, and other surveys, as well as the photometry of the candidate. For this event, all the sources were unrelated to the GRB.}
    \label{fig:grb221110}
\end{figure*}

As mentioned above, as a first demonstration of the utility of the platform, we perform an ``offline'' search for $\gamma$-ray burst afterglows from Fermi-GBM with ZTF using \texttt{fritz}. ZTF has employed an ongoing program to search for $\gamma$-ray bursts \citep{CoAh2019,AhSi2021,AhAn2022} and gravitational-waves~\citep{CoAh2019b,AnCo2020} using ``triggered'' Target of Opportunity (ToO) observations, which use timing and/or localization information from other wavelengths or messengers. It is similarly possible to use ``serendipitous'' observations akin to those performed within routine survey observations, which have neither localization nor explosion time information, to find counterparts as well \citep{AnKo2020,HoPe2020,AnCo2021}.

We use the short duration GRB\,200826A \citep{AhSi2021} to recreate the offline analysis of afterglows. For this, we download the Fermi XML VOEvent files and post them to \texttt{fritz}. Later, we trigger a customized query to post candidates within the days following an event to \texttt{fritz}. The spatial cross-match employed within PostgreSQL (sec.\,\ref{sec:hpa}) is used to keep candidates within the localization region. Limiting our queries to extragalactic fast transients, we keep transients at high Galactic latitude ($|b_{\rm Gal}| > 10$\,deg) with a fading rate faster than 0.3\,mag\,day$^{-1}$ in at least one band \citep{AnCo2021}. The results of our query are displayed in the localization map, showing ZTF20abwysqy among the sources recovered.

The first real-time ToO search using \texttt{fritz} was in November of 2022. We observed the localization region of the long GRB\,221110A (trigger 689784662) detected by the Gamma-Ray Burst Monitor (GBM) on the Fermi satellite with ZTF. We obtained a series of $g$- and $r$-band images covering 470 square degrees beginning at 09:47 UT on 2022 Nov. 11 ($\sim$ 15\,hr after the burst trigger time). This corresponded to $\sim$\,62\% of the probability enclosed in the Earth-occultation corrected GRB localization map. Each exposure was 240\,sec, reaching $g$- and $r$-band median depths of 21.2\,mag and 21.1\,mag respectively. The images were processed in real-time through the ZTF reduction and image subtraction pipelines at IPAC \citep{Masci2019}.

We queried the ZTF alert stream using Kowalski \citep{Duev_2021} and AMPEL \cite{Nordin:2019kxt}. We required at least 2 detections separated by at least 15 minutes to select against moving objects. The candidates within the 95\% probability contour of the GRB localization map discovered for the first time after the GRB trigger time, and that have more than 2 detections are displayed in Fig.~\ref{fig:grb221110}.

Further analysis consists of cross-matching our candidates with the Minor Planet Center to flag known asteroids, reject stellar sources \citep{TaMi2018}, and apply machine learning algorithms \citep{Mahabal2018}. These selection criteria as well as human vetting left us only with one candidate: ZTF22abtfyhi/AT2022aaaa. This transient was discovered at $g$ = 20.0\,mag, but later found to be rising (and therefore uninteresting) after SEDM optical follow-up.

\begin{figure}[t]
    \centering
    \includegraphics[width=3.5in]{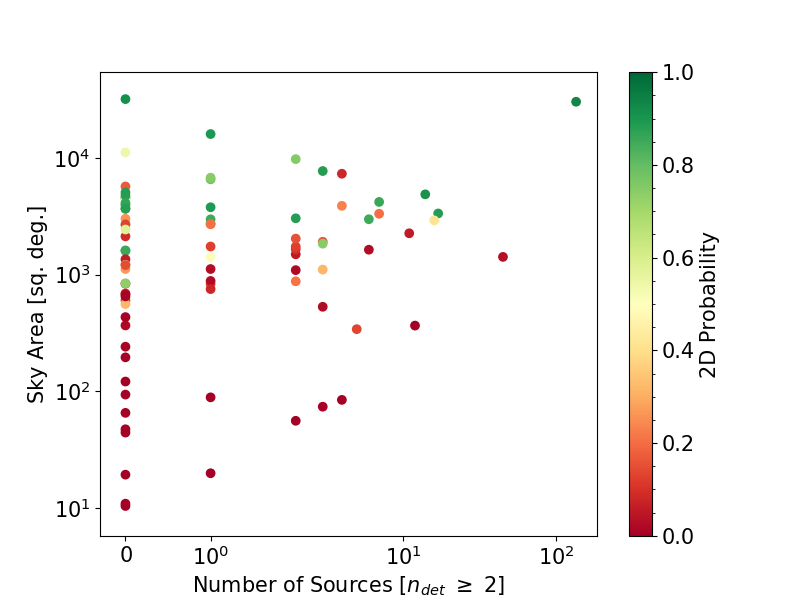}
    \caption{Scatter plot of sky area in square degrees vs. number of candidates (with at least 2 detections). The colors correspond to probability.}
    \label{fig:coverage}
\end{figure}

Reperforming this check across a variety of serendipitous observations of $\gamma$-ray burst skymaps yields candidate quantities shown in Fig.~\ref{fig:coverage}.

\section{Development Ethos and Conclusions}
\label{sec:conclusion}

In this paper, we have described the current state of multi-messenger functionalities within \texttt{SkyPortal} and potential future functionality. The rapid approach of O4 will provide an opportunity to identify additional needs for enabling these tools to allow near-automated identification, follow-up, and characterization of potential counterparts.
As an easily extensible framework, we plan a variety of improvements to the application; for example, we plan to be the first adopter of M4OPT, the Multi-Mission Multi-Messenger Observation Planning Toolkit package\footnote{\url{https://github.com/m4opt/m4opt}}, which is inspired by a previously prototyped Integer Linear Programming (ILP) based schedule for ZTF \citep{PaSi2022}.

We have described herein a number of frontend-focused workflows, recognizing that science in the MMA era, while greatly enhanced by software and analysis tools, remains fundamentally a collective human endeavor. Recognizing that many workflows also happen programmatically outside of \texttt{SkyPortal} the major functionality also is accessible via a well-documented API infrastructure\footnote{\url{https://skyportal.io/docs/api.html}}. It will come as no surprise that working across teams in different institutions---sometimes with aligned and sometimes with competing incentives---requires modern frameworks like \texttt{SkyPortal} to enable the establishment and curation of strict data access controls. Every photometry, spectroscopy, comment, annotation, classification entry in \texttt{SkyPortal} is permissioned by group.

\texttt{SkyPortal} is, first and foremost, scientific software developed to enable astronomical data science. However, a close secondary goal is to train students in proper software development practices, which has been identified as a gap between coursework and eventual readiness for academic and industry practice \citep{LiGa2022}. These required skills include software engineering skills (e.g., identifying requirements, prototyping, designing, and using version control) as well as transferable and soft skills (e.g., teamwork, and communication).

\texttt{SkyPortal} has had significant contributions from students at a variety of levels, with the faculty / permanent researchers associated with the project focused on bringing in students into the fold. For example, two sets of ten students from the Leonard de Vinci Engineering School have participated in the development of the features, focusing on the \texttt{icare} version of the application. This particular effort has yielded multiple internships at member institutions for these students.

However, bringing on students continues to raise a number of challenges for a small project like  \texttt{SkyPortal}. First of all, the onboarding efforts for an application the size of \texttt{SkyPortal} can be challenging. For this reason, the documentation is focused on both starting the application, making basic changes, and then an example of creating a pull request is quite thorough and continues to be updated as students join the project. However, significant dedicated senior developer time is still required not only initially during the onboarding process as well as later on as the students have become comfortable contributing; domain knowledge and project scope, as well as addressing user requests, still require significant input. This is potentially problematic given that these academic projects are led by professors with many other demands on their time.

Additionally, the international nature of this project puts an added strain on these dynamics. In addition to natural challenges with language barriers, finding times when the entire team can meet is very challenging, and in practice, multiple meetings per week with limited overlaps in persons due to time zone differences creates some challenges with ensuring a cohesive project direction. 

This is further challenged by active interactions with users, which occurs in a variety of venues; as for any project, features and bug fixes are done in conjunction with user requests, which are important for sustaining both the health of the application (for reporting bugs that are not caught in the Unit Testing) as well as needed features. These user interactions occur in a variety of ways, including through a mix of Announcements through GitHub discussions, e-mail updates, Google forms, Slack and other means of communication. This level of flexibility is convenient for users, but can be challenging for developers to track feedback in a variety of forms, not only the form the communication takes but also the level of feedback (i.e.\, small bug fixes vs.\ large feature requests).

In this way, \texttt{SkyPortal} can serve as an important scientific platform for this new era of multi-messenger astronomy. We expect that software platforms of this type will become prevalent to meet the needs of the scientific user community. We expect many projects will face similar challenges as their user communities grow and students contribute; we hope that the introduction of these projects for students and others will continue to serve as important opportunities to do science while also learning best practices in a technical environment.

\begin{acknowledgments}
The authors appreciate comments from Dave Coulter, Fabian Schussler, and an anomymous referee on an initial draft of this paper.

MWC, JL, and VS are supported by the National Science Foundation with grant number OAC-2117997; MWC is also supported by PHY-2010970. 
MWC and RWK were supported by the Preparing for Astrophysics with LSST Program, funded by the Heising Simons Foundation through grant 2021-2975, and administered by Las Cumbres Observatory.
The Gordon and Betty Moore Foundation, through both the Data-Driven Investigator Program and a dedicated grant to develop \texttt{SkyPortal}, provided critical funding for this project without which this project could not have succeeded. J.S.B., G.N., A.C-Q., D.A.D.\ were partially supported by the Gordon and Betty Moore Foundation. BP thanks the Northeastern Lawrence Co-op Fellowship for their continuous support. JP thanks LSST-France and CNRS/IN2P3 for supporting Fink.
The Kilonova-Catcher program is supported by the IdEx Université de Paris Cité, ANR-18-IDEX-0001 and the MITI CNRS Sciences participative program.

Based on observations obtained with the Samuel Oschin Telescope 48-inch and the 60-inch Telescope at the Palomar Observatory as part of the Zwicky Transient Facility project. ZTF is supported by the National Science Foundation under Grant No. AST-2034437 and a collaboration including Caltech, IPAC, the Weizmann Institute of Science, the Oskar Klein Center at Stockholm University, the University of Maryland, Deutsches Elektronen-Synchrotron and Humboldt University, the TANGO Consortium of Taiwan, the University of Wisconsin at Milwaukee, Trinity College Dublin, Lawrence Livermore National Laboratories, IN2P3, University of Warwick, Ruhr University Bochum and Northwestern University. Operations are conducted by COO, IPAC, and UW.

\end{acknowledgments}



\bibliographystyle{aasjournal}
\bibliography{references} 

\label{lastpage}
\end{document}